\documentclass[
superscriptaddress,
preprint,
bibnotes,
amsmath,amssymb,
aps,
prmaterials,
]{revtex4-2}

\usepackage{graphicx}% Include figure files
\usepackage{dcolumn}% Align table columns on decimal point
\usepackage{bm}% bold math
\usepackage{booktabs} 
\usepackage{amsmath} 
\usepackage[utf8]{inputenc}
\usepackage[hidelinks]{hyperref}% add hypertext capabilities
\usepackage{float}
\usepackage{subcaption}
\usepackage{xcolor}
\usepackage{fitbox}
\usepackage{caption}
\makeatletter
\newcommand*{\rom}[1]{\expandafter\@slowromancap\romannumeral #1@}
\usepackage{adjustbox}
\DeclareUnicodeCharacter{2212}{\ensuremath{-}} %ask Edward for better way of solving this
\begin{document}

\title{Predicting the suitability of photocatalysts for water splitting using Koopmans spectral functionals: The case of TiO\textsubscript{2} polymorphs} %Force line breaks with \\

\author{Marija Stojkovic}
\email{marija.stojkovic@epfl.ch}
\affiliation{Theory and Simulations of Materials (THEOS), \'{E}cole Polytechnique F\'{e}d\'{e}rale de Lausanne, 1015 Lausanne, Switzerland}
\author{Edward Linscott}
\affiliation{Center for Scientific Computing, Theory and Data, Paul Scherrer Institute, 5232 Villigen PSI, Switzerland}
\affiliation{National Centre for Computational Design and Discovery of Novel Materials (MARVEL), Paul Scherrer Institute, 5232 Villigen PSI, Switzerland}

\author{Nicola Marzari}
\email{nicola.marzari@epfl.ch}
\affiliation{Theory and Simulations of Materials (THEOS), \'{E}cole Polytechnique F\'{e}d\'{e}rale de Lausanne, 1015 Lausanne, Switzerland}
\affiliation{Center for Scientific Computing, Theory and Data, Paul Scherrer Institute, 5232 Villigen PSI, Switzerland}
\affiliation{National Centre for Computational Design and Discovery of Novel Materials (MARVEL), Paul Scherrer Institute, 5232 Villigen PSI, Switzerland}

\date{\today}% It is always \today, today,
\begin{abstract}
Photocatalytic water splitting has attracted considerable attention for renewable energy production. Since the first reported photocatalytic water splitting by titanium dioxide, this material remains one of the most promising photocatalysts, due to its suitable band gap and band-edge positions. However, predicting both of these properties is a challenging task for existing computational methods. Here we show how Koopmans spectral functionals can accurately predict the band structure and level alignment of rutile, anatase, and brookite TiO\textsubscript{2} using a computationally efficient workflow that only requires (a) a DFT calculation of the photocatalyst/vacuum interface and (b) a Koopmans spectral functional calculation of the bulk photocatalyst. The success of this approach for TiO\textsubscript{2} suggests that this strategy could be deployed for assessing the suitability of novel photocatalyst candidates. 
%\begin{description}
%\end{description}
\end{abstract}

\maketitle

%\tableofcontents

\section{\label{sec:level1}Introduction}

One of the most pressing problems that we are currently facing is finding easy and low-cost renewable energy sources. Hydrogen production from water is one attractive option. In this process, water is decomposed by visible light into oxygen and hydrogen without the application of external potentials, as was first demonstrated by Fujishima and Honda using $\mathrm{TiO_2}$ as an electrode \cite{FujishimaandHonda1972}. Ever since, photocatalytic water splitting (PWS) has been in the spotlight as a way to produce hydrogen via renewable energy. Alongside experiments that have explored PWS at a fundamental level --- from photon absorption to the production of molecular hydrogen --- computational methods have aided our understanding of this process, and are particularly useful for identifying new candidate materials for catalysts. The search for an ideal photocatalytic material is still ongoing, and numerous studies have been conducted to tackle this problem, especially for semiconductor materials (see \cite{kudo2009heterogeneous, ni2007review, maeda2010photocatalytic, takanabe2017photocatalytic, yang2014semiconductor,fu2018material} and references therein).

\begin{figure}
    \centering
    \includegraphics[width=\columnwidth]{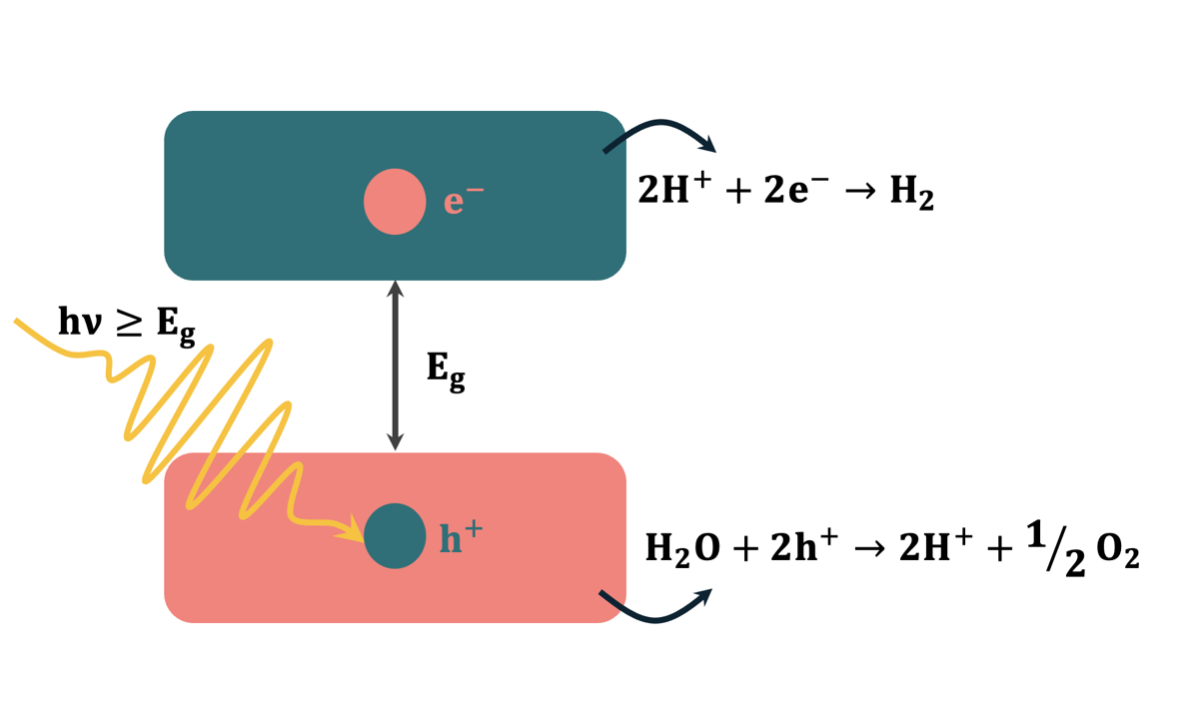} 
    \caption{Schematic illustration of photocatalytic water splitting. The conduction-band and valence-band regions are shown in green and red, respectively. When a photon (yellow line) with energy equal to or greater than the semiconductor band gap is absorbed, an electron is excited from the valence band to the conduction band, leaving a hole behind. The electron and hole subsequently participate in the reduction of hydrogen (top right) and oxidation of water (bottom right).}
    \label{fig:PCWS} 
\end{figure}

To better understand the desirable properties for a candidate photocatalyst, let us first briefly revise the PWS process. The process is schematically presented in Figure \ref{fig:PCWS}.
Water splitting is, of course, not a spontaneous reaction: it requires an energy barrier to be overcome. One of the ways to do this is via solar irradiation. Upon light absorption, the catalyst plays a crucial role in a three-step process: 1.\ charge carrier generation, in which electrons are promoted from the valence to the conduction band (creating a hole); 2.\ charge separation of the electron-hole pair and migration to the surface of a catalyst; and 3.\ participation in redox reactions. The first two steps are especially dependent on the structural and electronic properties of the photocatalytic material. The overall process is given by
\begin{equation}
 \begin{split}
  \mathrm{ 2H_2O (l) \longrightarrow O_2(g) + 4H^+ + 4e^-
  \hspace{5mm}E_{ox}^o = 1.23V}
 \end{split}
 \label{eqn:H2O_ox}
\end{equation}
 
\begin{equation}
\mathrm{
    4H^+ + 4e^- \longrightarrow 2H_2(g) \hspace{5mm} E_{re}^o = 0.00V}
    \label{eqn:H_red}
\end{equation}

%\begin{equation}
%\mathrm{
%   2H_2O \longrightarrow 2H_2(g) + O_2(g) \hspace{5mm}E_{cell}^o = %    \label{eqn:overall}
%\end{equation}

The water-splitting reaction involves the oxidation of water and reduction of hydrogen. 
Under standard conditions the value of the redox potentials --- as referenced to the normal hydrogen electrode --- are 1.23 V and 0 V respectively \cite{CRCHandbook}. 
Given these potentials, the band gap of a suitable catalyst needs to be at least 1.23 eV; otherwise the electrons will not have enough energy to start the reaction. In practice, at least 1.6 to 1.8 eV are required, as a certain amount of excess energy (\emph{i.e.} the so-called ``overpotential") is needed to overcome kinetic barriers and induce the hydrogen and oxygen evolution reactions on the surface of the electrode \cite{jafari2016photocatalytic, walter2010solar, park2019mechanism}. The band gap should not be too high either, since that would reduce the amount of visible light that the photocatalyst can harness.

However, an optimal band gap is not enough. For the PWS reaction to occur, the valence band maximum (VBM) needs to be higher than the oxidation potential of water while the conduction band minimum (CBM) needs to be lower than the hydrogen reduction potential. This ensures that the overall reaction has a negative change in Gibbs free energy, and is therefore spontaneous.

%And while this is true as well for two half redox reaction during photocatalysis, the net Gibbs free energy can be both positive and negative. However, in order to achieve spontaneous half redox reactions, they need to be spatially or chemically separated (for both $\Delta G<0$ and $\Delta G > 0$), and in comparison with their standard electrode potentials CBM and VBM need to be at specified positions on a potential scale. Therefore in some cases, even with the appropriate band gap, if the VBM and CBM are not properly aligned then the PWS reaction will not occur.

% Placing this here so that it appears at the top of the second page
\begin{figure*}[t]
\centering

\renewcommand{\arraystretch}{0.8} % Adjust row height for better spacing

\begin{tabular}{p{0.25\textwidth} p{0.4\textwidth} p{0.25\textwidth} }
    \includegraphics[width=\linewidth]{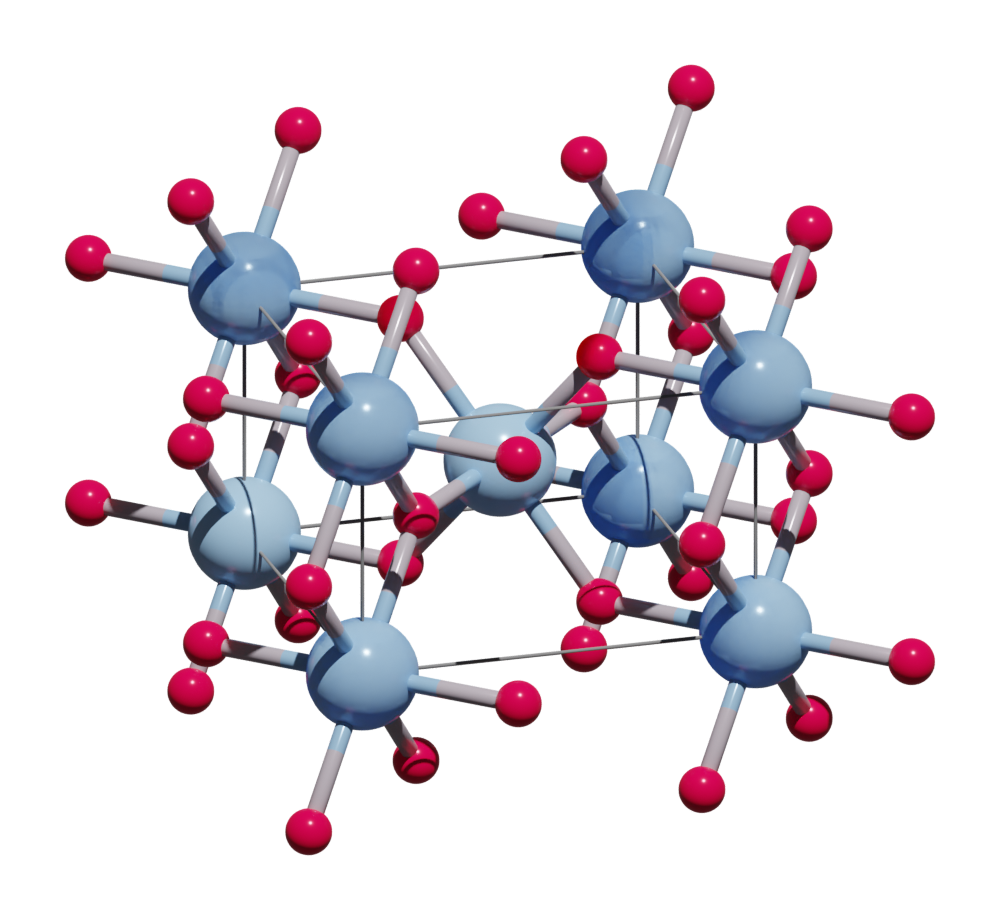} & 
    \includegraphics[width=\linewidth]{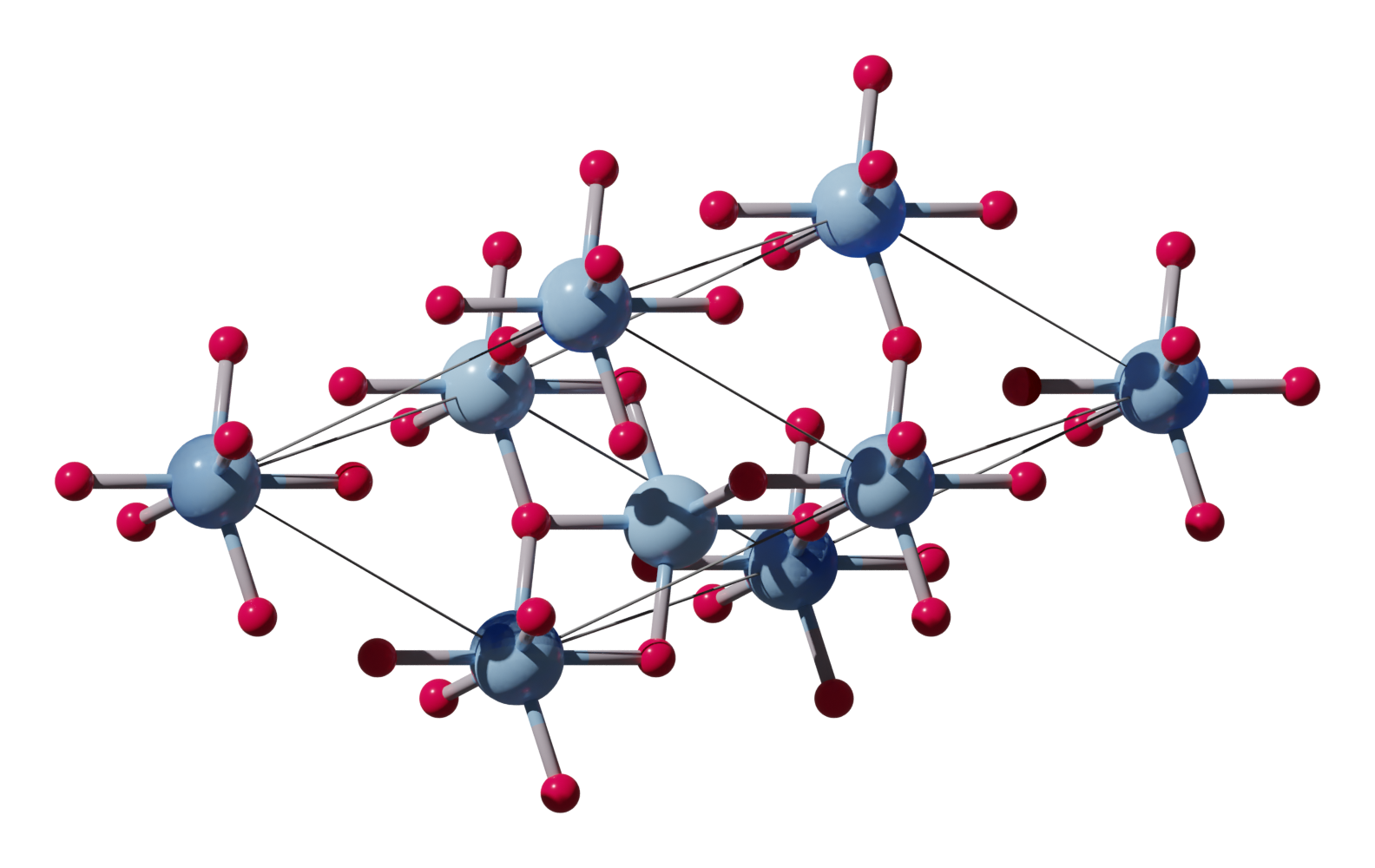} & 
    \includegraphics[width=\linewidth]{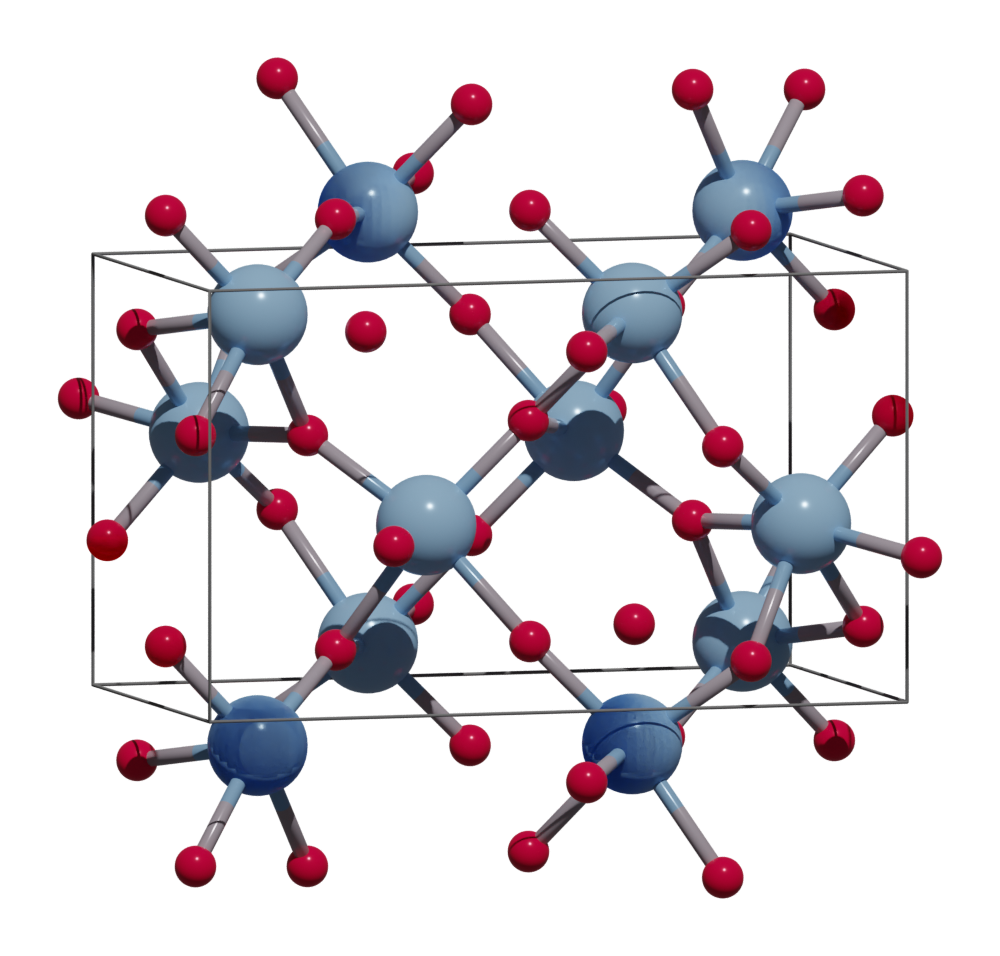} \\ 
    (a) rutile 
    & (b) anatase
    & (c) brookite \\
    \scriptsize 6 atoms/cell &
    \scriptsize 6 atoms/cell &
    \scriptsize 24 atoms/cell \\
    \scriptsize primitive tetragonal &
    \scriptsize body-centered tetragonal & 
    \scriptsize primitive orthorhombic \\
    \scriptsize $P4_2/mnm$ & 
    \scriptsize $I4_1/amd$ & 
    \scriptsize $Pbca$ \\
    \scriptsize $a = b = 4.6$, $c = 2.96$\,\AA & 
    \scriptsize $a = b = 3.80$, $c = 9.68$\,\AA & 
    \scriptsize $a = 5.14$, $b = 9.25$, $c = 5.50$\,\AA \\
\end{tabular}
\caption{Crystal structures of three $\mathrm{TiO_2}$ polymorphs}
\label{fig:figures}
\end{figure*}

% \begin{figure*}[t]
% \centering
% \begin{subfigure}[t]{0.25\textwidth}
%     \includegraphics[width=\textwidth]{Figures/Rutile_bulk_primitive.png}
%     \caption{rutile
%     \begin{minipage}[t]{\linewidth}
%         primitive tetragonal (P4\textsubscript{2}/mnm) \\ $a=b= 4.6$, $c= 2.96$\,\AA
%     \end{minipage}
%     }
%     \label{fig:first}
% \end{subfigure}
% \begin{subfigure}[t]{0.4\textwidth}
%     \includegraphics[width=\textwidth]{Figures/Anatase_bulk_primitive_rotated.png}
%     \caption{anatase
%     \begin{minipage}[t]{\linewidth}
%         body-centered tetragonal (I4\textsubscript{1}/amd) \\ $a=b=3.80$, $c=9.68$\,\AA
%     \end{minipage}
%     }
%     \label{fig:second}
% \end{subfigure}
% \begin{subfigure}[t]{0.25\textwidth}
%     \includegraphics[width=\textwidth]{Figures/Brookite_bulk_primitive.png}
%     \caption{brookite
%     \begin{minipage}[t]{\linewidth}
%         primitive orthorhombic (Pbca) \\ $a=5.14$, $b=9.25$, $c=5.50$\,\AA
%     \end{minipage}
%     }
%     \label{fig:third}
% \end{subfigure}
% \caption{Crystal structures of three $\mathrm{TiO_2}$ polymorphs}
% \label{fig:figures}
% \end{figure*}

Given the importance of the band gap and the band alignment on the performance of photocatalysts, we would like to be able to computationally predict these properties. However, this is a challenging task.  In computational materials science, Kohn-Sham density functional theory (KS-DFT) is ubiquitous, but its prediction of band gaps and band alignment is unreliable. This failure stems from several shortcomings. Firstly, the Kohn-Sham eigenvalues (from which we obtain both the band gap and band alignment) are actually mathematical construct and do not have a direct physical meaning --- they need not correspond to genuine excitation energies of the system \cite{PhysRevB.60.4545,PhysRevLett.112.096401}. This is true even for the exact exchange-correlation (xc) functional, for which only the highest occupied molecular orbital (HOMO) has an energy that corresponds to a physical excitation. Approximate xc functionals additionally suffer from further errors such as ``one-body" self-interaction error (the incomplete cancellation of the Hartree and exchange terms for one-electron systems) \cite{PhysRevB.23.5048, PhysRevLett.97.103001} and an erroneous curvature of the total energy with respect to the total number of particles (which should in principle be piecewise linear) \cite{{perdew1982density,cohen2012challenges,mori2008localization,D0CP02564J,dabo2010koopmans,PhysRevB.71.035105}}. The lack of piecewise linearity means that the eigenvalues obtained via approximate xc functionals do not match the total energy differences that one would obtain by explicitly performing electron removal/addition \cite{perdew1982density,harbola1999relationship,dabo2010koopmans}.

%Firstly, semi-local (DFT) energies  exhibit erroneous curvature with respect to the total number of particles instead of being piecewise linear (PWL) \cite{perdew1982density,cohen2012challenges,mori2008localization}
%Missing PWL , in consequence, makes interpretation of Kohn-Sham eigenvalues difficult, meaning that we cannot obtain the exact values of ionization potential as defined in IP theorem.\cite{perdew1982density,harbola1999relationship}. Ultimately, Kohn-Sham eigenvalues are a mathematical construct and --within the exception of the HOMO --need not correspond to physically-- meaningful excitation energies\cite{PhysRevB.60.4545,PhysRevLett.112.096401}. Finally, semi-local exchange correlation functional do not precisely cancel the self-Hartree term\cite{PhysRevB.23.5048}.

Higher-order methods have been used to overcome these shortcomings. These include hybrid functionals \cite{heyd2003hybrid,marsman2008hybrid,perdew1982density,becke1993half, kraisler2020asymptotic}, Hubbard and extended Hubbard functionals \cite{himmetoglu2014hubbard,campo2010extended}, and many-body perturbation theories such as GW \cite{aryasetiawan1998gw,shishkin2006implementation,golze2019gw, PhysRevB.84.193304}. The first two approaches (partially) address the issue of piecewise linearity, but still do not have eigenvalues that can be formally interpreted as quasiparticle energies. GW, on the other hand, has formally-well-defined eigenvalues (being a theoretical framework centered on the description of quasiparticles) but (a) it is much more computationally demanding and (b) performing self-consistency does not systematically improve the results --- for example, self-consistent quasiparticle GW typically overestimates band gaps and in this regard is less accurate than G\textsubscript{0}W\textsubscript{0} \cite{doi:10.1021/ct500087v}.

In this work we employ Koopmans spectral functionals \cite{dabo2010koopmans} as an alternative approach that provides accurate spectral properties of materials while being less computationally demanding compared to many-body approaches. As a test case, we apply these functionals to the three most stable polymorphs of TiO\textsubscript{2}: rutile, anatase, and the less-studied brookite \cite{hiroi2022inorganic}. The crystal structures of these three polymorphs are presented in Figure~\ref{fig:figures}. %Although TiO\textsubscript{2} is very well-known, studies of these materials still seem to be sparse, in particular for the brookite phase.% In this paper we obtain the band gap and band level alignment for three polymorphs with good agreement with experimental studies.

The paper is organized as follows: in Section \ref{sec:Theoretical framework} we explain the theoretical framework behind band alignment and how these calculations need to be adapted for Koopmans functionals. In Section \ref{Method} we describe the computational details and methodology that we use. Finally, in Section \ref{Results and discussion} we present and discuss the electronic structure calculations of the three polymorphs and their band alignment via Koopmans functionals, and compare the results against other existing methods.

\section{Theoretical framework}\label{sec:Theoretical framework}
\noindent As has already been mentioned, the valence and conduction band positions with respect to the water redox levels are some of the most important features of a promising photocatalyst. The positions of the band edges depend on --- among other factors --- the crystal structure, the chemical environment, and the surface structure of the material \cite{park2019optimal,scanlon2013band,pfeifer2013energy}. 
%what is an interface for us?
Broadly speaking, ``band alignment" refers to measuring the relative alignment of the energy bands of two materials at an interface. Often this interface is between two solid materials (such as semiconductors) but it can also refer to a junction between a semiconductor and fluid (such as the case of water splitting via photocatalytic surface).
% how do we determine band alignment(general categorization)
 There are several methods for calculating the alignment of bands at an interface, which can be broadly categorized based on the choice of reference level, or, relatedly, what system is modeled (e.g. the bulk alone \emph{vs.} the explicit interface).
  % Band alignment --obtained by modeling a bulk system
The band alignment can be estimated from the bulk structure alone via branch point energies (BPEs) \cite{PhysRevLett.52.465}. BPEs are defined as the energy points where bands change character from being predominantly donor-like to acceptor-like. It has been shown that the electronegativity and variation in character of interface-induced gap states are determined by the nearest band edge, which can in turn be used to estimate the band offset \cite{MONCH1997380}. One clear advantage of this method is that it only requires calculations of the bulk, drastically reducing its computational cost, but the method cannot be applied to materials for which the BPEs lie in the conduction band region \cite{schleife2009branch}.
 % Band alignment -- modeling heterointerfaces
 % Furthermore a method that relies solely on information about the bulk alone can not provide a complete description of the electron distribution at the interface. In order to obrtain complete picture i.e. to fully describe the electron distribution at the interface explicit modeling of hetero-interfaces is used. Generally this method can be applied on hetero-interfaces with and without the lattice missmatch.If we consider lattice matched systes the junction between two solids shift in the average potentials occurs, which can be taken as the reference with respect bulk band structures can be aligned.  
 At the other end of the spectrum, the hetero-interface can be explicitly modeled. For hetero-interfaces without lattice mismatch, this approach is straightforward because it only requires aligning the valence and conduction bands of semiconductors against the calculated average potentials in the plane parallel to the interfaces \cite{PhysRevB.35.8154}. In the case of lattice mismatch, potential deformation of a core state needs to be taken into account \cite{PhysRevB.73.245206, PhysRevB.50.17797}. 
 % Band alignment -- modeling slab structures
Finally, an intermediate approach is to measure the band offsets with respect to a reference level. This reference level can be the vacuum level, in which one obtains the ionization potential (IP) and electron affinity (EA) of each material \cite{stevanovic2014assessing}. These properties are intrinsically surface properties and must be obtained by considering a slab model.
% Band alignment using a different reference level
Alternatively, an impurity can be used as a reference level, such as hydrogen \cite{van2003universal}
 or transition metal impurities \cite{PhysRevB.38.7723}, where band alignment relies on the position of dangling bonds formed between the impurity and semiconductor. 
%Other methods
Other more approximate approaches exist, such as effective dipole moments, tight-binding schemes, and empirical schemes \cite{nano11061581,PhysRevB.30.4874}. All of the above methods are nicely summarized in  Ref.~\onlinecite{hinuma2014band}. In this work, we use the method based on the IP and EA of the material, adapted for use with orbital-density-dependent functionals \cite{kang2010quasiparticle,yin2010band,shaltaf2008band,wu2011prediction}.
 
\subsection{Band alignment with DFT}
\label{sec:dft_band_alignment}
\noindent Before discussing how band alignment calculations with respect to the vacuum are performed for orbital-dependent functional theories, we first briefly review how these calculations are performed for standard functionals.

\begin{figure}[t]
    \centering
    \includegraphics[width=\columnwidth]{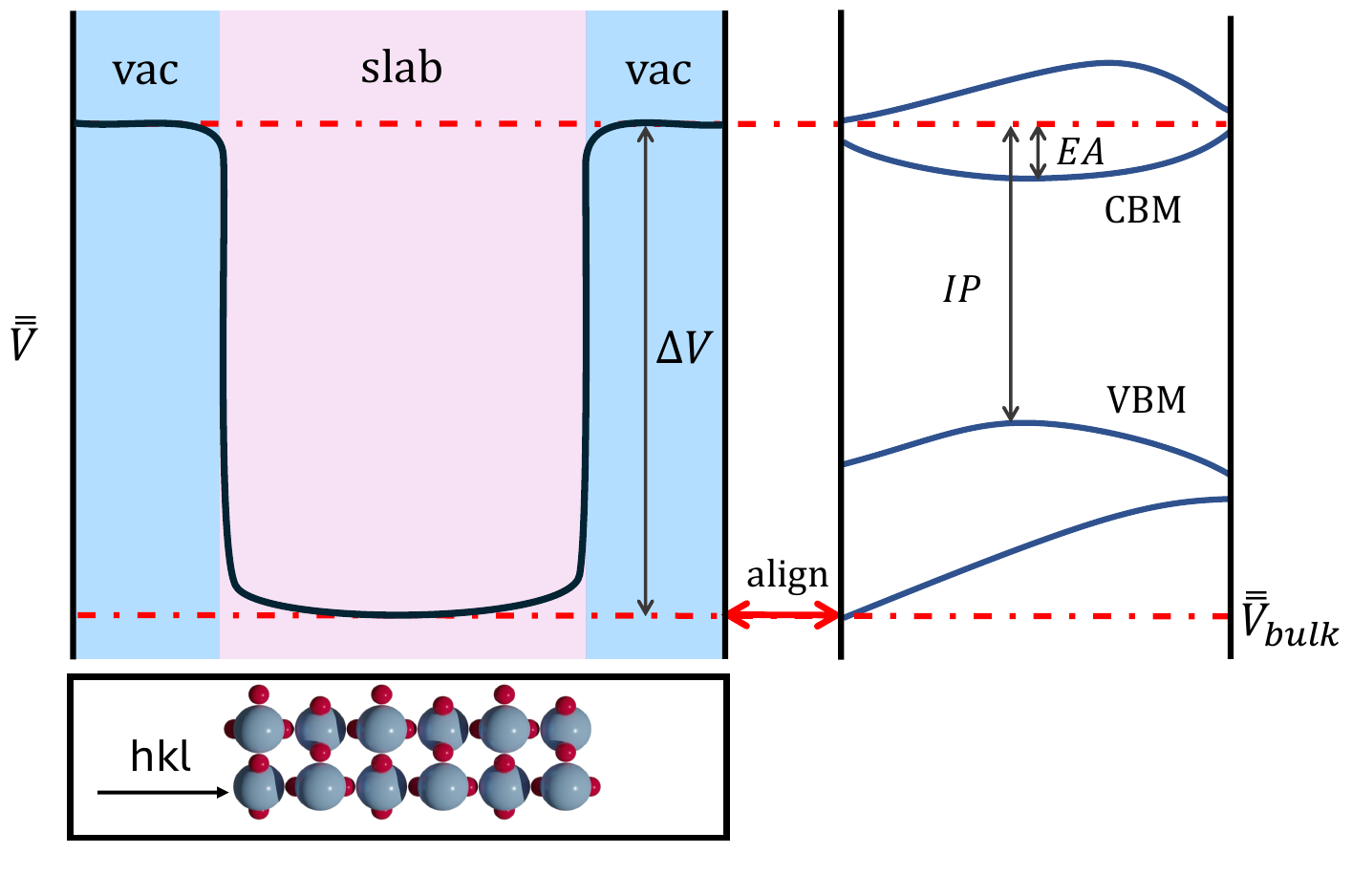}
    \caption{Cartoon of the band alignment procedure. The black line represents the macroscopic average potential $\Delta V$ calculated across the slab for most stable surface facet. The IP and EA correspond to the VBM and CBM relative to the vacuum reference level, which is obtained via alignment of the average potential $V_\mathrm{bulk}$ for the slab and bulk systems.}
    \label{fig:band_alignment}
\end{figure}

In order to calculate the ionization potential and electron affinity of a material, we must be able to reference the valence band maximum (VBM) and conduction band maximum (CBM) against the vacuum level. However, the VBM and CBM are extracted from DFT calculations as Kohn-Sham eigenvalues, whose absolute value carry no physical meaning: indeed, the Kohn-Sham potential can be shifted by an arbitrary constant leaving the physical system unchanged but shifting the KS eigenvalues by the same amount. To tether the KS eigenvalues to something meaningful, one valid choice is the potential of the vacuum. To obtain the potential in the vacuum, one considers a slab model, where the bulk catalyst is interfaced with vacuum, and then calculates the change in the average electrostatic potential between the slab and vacuum regions (denoted $\Delta V$). This scheme is illustrated in Figure \ref{fig:band_alignment}.

The average electrostatic potential is obtained by first calculating a planar average of the three-dimensional electrostatic potential: 
\begin{equation}
     \overline{V}(z) = \frac{1}{S}\int_S dxdy V(r) 
 \end{equation}
 where $S$ is the cross-sectional area of the cell parallel to the interface and for simplicity we have assumed a tetragonal cell with a slab oriented in the $z$ direction. Within the slab region, $\overline{V}(z)$ exhibits oscillations with a periodicity matching that of the ionic cores. These oscillations can be removed by macroscopically averaging the potential:
 \begin{equation}
\overline{\overline{V}}(z) = \frac{1}{L} \int_{z-\frac{L}{2}}^{z+\frac{L}{2}} \overline{V}(z')dz'
 \end{equation}
using an averaging window of length $L$ that matches the periodicity of the lattice.

Having simulated a slab model and extracted $\Delta V$, the vacuum level for a bulk system is then given by the average electrostatic potential across the entire cell $\langle V\rangle_\mathrm{bulk}$ plus the potential difference $\Delta V$ as calculated from the slab calculation. The IP and EA are then given as the (negative of the) VBM and CBM energy levels relative to this vacuum level \emph{i.e.}
\begin{equation}
    \label{eqn: ip_dft}
    \mathrm{IP} = \Delta V - \varepsilon_\mathrm{VBM}
\end{equation}
and
\begin{equation}
\label{eqn:ea_dft}
    \mathrm{EA} = \Delta V - \varepsilon_\mathrm{CBM}
\end{equation}
We stress that this procedure relies on (a) a sufficiently thick slab, so that the electrostatic potential deep within the slab is bulk-like \cite{van1987theoretical}, and (b) a sufficiently wide vacuum region, in order for the electrostatic potential to converge. It is also important to allow for structural relaxation of the slab/vacuum interface as this can significantly affect the potential difference $\Delta V$ \cite{weston2018accurate}.

%It is known that band alignment of semiconductors and insulators can be obtained solely from their ionization potentials (IPs) and electron affinities (EAs) \cite{stevanovic2014assessing}. This stems from the fact that IPs and EAs are the measure of the position of the valence band maximum (VBM) and conduction band minimum (CBM) respectively. More precisely IP is defined as a difference between the reference level and the highest occupied level in the bulk region, or in some cases it can be referenced to the surface levels\cite{hinuma2012ionization}. However, determining the IP imposes a question: How do we choose the reference level since the total energy that we obtain from the SCF calculation does not give us any actual physical meaning?
%This implies that we are obliged to reference the eigenvalues against some other energy level. For this purpose vacuum level is usually used, and fundamental reason for this lies in the long-range Coulomb interaction\cite{van1987theoretical} where the charge density on the surface determines the positions of the energy levels deep inside the bulk. Some claim that impurity can also play that role, and this is usually used in hetero-junction systems\cite{robertson2013band}. To illustrate this we refer to the Figure \ref{fig:band_alignment}.

This approach has one notable drawback: semi-local DFT eigenvalues are often not quantitatively (nor even qualitatively) accurate. It is therefore necessary to go beyond DFT to obtain accurate predictions of the valence and conduction band positions.

\subsection{Koopmans spectral functionals}
% In an attempt to accurately describe the spectral properties of materials, Koopmans functionals emerged .
%These functionals are based on two key properties. Firstly following Janak's theorem\cite{PhysRevB.18.7165} the orbitals eigenvalues are independent of its occupations and the second one is that they are proportional to the energy difference corresponding to the electron addition and removal(energy difference between integer number of particles). The first property removes curvature for each orbital while the second property exactly imposes the PWL described earlier.
%In this case Koopmans specral functionals impose so called general piecewise-linearity (GPWL) which states that...

\noindent But why are KS-DFT eigenvalues unreliable, and what can we do to improve them? To start to answer this question, we note that DFT Kohn-Sham eigenvalues do not match the corresponding energy difference one obtains from explicitly performing electron addition/removal (\emph{i.e.}\ a $\Delta$SCF calculation). Contrast this with the exact one-body Green's function --- an object that describes one-particle excitations \emph{exactly} --- whose excitation energies (\emph{i.e.}\ poles) are located precisely at points that correspond to addition/removal energies (as can be straightforwardly seen from its spectral representation).

Inspired by this observation, Koopmans functionals are a class of functionals that seek to accurately describe the spectral properties of materials by restoring eigenvalue/total-energy-difference equivalence to DFT \cite{dabo2010koopmans}. To do so, they impose the so-called ``generalized piecewise-linearity'' (GPWL) condition, which states that the orbital energies $\varepsilon_i$ of orbitals $\varphi_i$ should be independent of that orbital's occupation $\emph{f}_i$:
\begin{equation}
    \varepsilon_i = \langle\varphi_i|\hat{H}|\varphi_i\rangle = \frac{dE}{d\emph{f}_i} 
    \label{eq:GPWL}
\end{equation} 
This is related to the more well-known ``piecewise linearity'' (PWL) condition i.e.\ linearity of the total energy with respect to the total number of electrons in the system. Standard density functional approximations are not piecewise-linear, causing the aforementioned discrepancy between total energy differences and eigenvalues.

The general form of a functional that imposes the GPWL condition is: 
\begin{equation}
    E_\mathrm{Koopmans} [\rho,{f_i}]= E^\mathrm{DFT}[\rho] + \sum_i \left(- \int_{0}^{f_i} \varepsilon_i(f)df + f_i\eta_i\right)
    \label{eq:Koopmans-functionals-general-form}
\end{equation}
As this equation shows, Koopmans functionals take the form of a correction to DFT: The first term on the right-hand side, $E^\mathrm{DFT}$, is the energy of a (typically semi-local) ``base" functional. The corrective terms then impose GPWL (equation~\ref{eq:GPWL}) by removing, orbital-by-orbital, all non-linear dependence of the total energy on the orbital occupancies (the first term inside the parentheses) and replacing it with a term that is linear with respect to the orbital occupation $\emph{f}_i$ (the second term inside the parentheses) \cite{nguyen2015first,nguyen2018koopmans,borghi2014koopmans,borghi2015variational}. Here $\eta_i$ is some constant that will be defined shortly.

An important feature of Koopmans functionals is that they are not density functionals: the energy depends not only on the total density of the system but also on the densities of individual orbitals, making this an ``orbital-density-dependent'' functional theory. This comes with several important consequences. In the framework of DFT, the energy is invariant with respect to unitary rotations of the occupied manifold, and this means that the same set of orbitals minimize the total energy and diagonalize the Hamiltonian. This is not the case for orbital-density dependent functionals, for which the ``variational" orbitals minimize the energy while the ``canonical" orbitals diagonalize the Hamiltonian constructed at the end of minimization procedure. (Note that this does \emph{not} imply that the functional is not variational. The loss of unitary invariance and the emergence of variational and canonical orbitals is a feature common to all orbital-density dependent functionals including \emph{e.g.}\ PZ-SIC.)

Let us now return to the $\eta_i$ term in equation~\ref{eq:Koopmans-functionals-general-form}. Depending on how one chooses $\eta_i$, different flavors of Koopmans can be defined \cite{nguyen2015first}. The first of these, the Koopmans integral correction (KI), sets $\eta_i$ to guarantee that the orbital energies $\varepsilon_i$ are equal to the corresponding $\Delta$SCF total energy difference of the base functional. Explicitly, the KI functional is given by

\begin{widetext}
\begin{equation}  
    E^\mathrm{KI}[\{\rho_i\}] = E^\mathrm{DFT}[\rho] + \sum_i \alpha_i \Big[\Big(E_\mathrm{Hxc}[\rho-\rho_i] - E_\mathrm{Hxc}[\rho]\Big) + \\ f_i\Big(E_\mathrm{Hxc}[\rho -\rho_i +n_i] - E_{Hxc}[\rho-\rho_i]\Big)\Big]
  \label{eqn:ki-functional}
\end{equation}
\end{widetext}
where $\rho_i$ is orbital density at filling $f_i$ and $n_i$ is the normalized orbital density i.e. $\rho_i(\mathbf{r}) = f_i n_i(\mathbf{r})$.  The parameters $\alpha_i$ are screening parameters, which account for the fact that the derivation of the KI correction only accounts for the explicit dependence of the DFT energy on the orbital occupancies in going from equation \ref{eq:Koopmans-functionals-general-form} to \ref{eqn:ki-functional}. By scaling the strength of the orbital correction via these screening parameters, we account for orbital relaxation \emph{post hoc}. Crucially, these screening parameters $\alpha_i$ are system-specific, and can be computed \emph{ab initio} --- \emph{i.e.} they are \emph{not} fitting parameters \cite{Colonna2018}.

Curiously,  for insulating systems where all the orbital occupancies are either 0 or 1, the ground state KI energies and densities match those of the base functional. (This can be seen by setting $f_i$ to 0 or 1 in equation~\ref{eqn:ki-functional}, in which case the entire correction vanishes. (We stress that this does not mean that the Koopmans correction has no effect, because the derivative of the correction is non-vanishing and thus the eigenvalues of the Koopmans functional will be different to those of the base functional.) This property of the KI functional allows us to obtain the KI ground-state density directly from a single semi-local DFT calculation, after which only unitary rotations of the occupied variational orbitals are needed to locate the KI minimum. This greatly reduces the computational cost of these calculations.

The second form of Koopmans functionals, KIPZ, adds a screened Perdew-Zunger (PZ) self-interaction correction term to the functional:
\begin{equation}
\label{eqn:kipz-functional}
    E^\mathrm{KIPZ} = E^\mathrm{KI} - \sum_i \alpha_if_iE_\mathrm{Hxc}[n_i]
\end{equation}
which removes one-body self-interaction error and guarantees that the functional is exact for one-electron systems. In contrast to KI, KIPZ does not share the same ground state density as the base functional and therefore mandates a full minimization of an orbital-density-dependent functional. Because of this, it can be advantageous to evaluate the KIPZ Hamiltonian on the KI ground state. This is referred to as perturbative KIPZ (pKIPZ).

Because Koopmans functionals are orbital-density-dependent and have screening parameters that must be calculated \emph{ab initio}, the procedure for computing a Koopmans band structure involves several steps as follows:
\begin{enumerate}
    \item a ground-state DFT calculation to initialize the total density
    \item a Wannierization of the DFT Kohn--Sham states in order to obtain maximally localized Wannier orbitals with which to initialize the variational orbitals
    \item calculating the screening parameters $\{\alpha_i\}$ via finite differences or density functional perturbation theory \cite{colonna2019koopmans,colonna2022koopmans}
    \item a final Koopmans calculation to minimize the energy and construct and diagonalize the Hamiltonian
\end{enumerate}
There are a few important technical details regarding this procedure. First, while any unitary rotation of the occupied manifold leaves the DFT energy unchanged, ODD functionals assign different energies to different sets of orbitals with the same total density. As a consequence, the minimization cannot rely on the usual self-consistent diagonalization procedure. Instead, the variational orbitals are obtained by directly minimizing the ODD energy using two nested loops, following the strategy introduced in ensemble DFT~\cite{marzari1997ensemble}: an inner loop optimizes the orbital rotations at fixed density, while an outer loop allows the density itself to relax. This conjugate-gradient procedure yields the self-consistent minimum of the functional (see Supplemental Material S1 \cite{SM} for more detail). Additionally, note that orbital occupations are not treated as variational parameters in this procedure. Since all systems considered here have an electronic gap, the occupied orbitals always have $f_i = 1$ and the empty orbitals $f_i = 0$, and only the orbitals themselves are optimized.

Second, the emergence of variational orbitals --- which are \emph{localized} orbitals --- is crucial because it allows Koopmans functionals to correctly treat bulk systems \cite{nguyen2018koopmans}. The Koopmans correction is applied to each variational orbital, while the canonical orbitals are interpreted as Dyson orbitals and their eigenvalues as quasiparticle energies, which are shifted by some amount depending on the size of the Koopmans corrections applied to each of their constituent variational orbitals. In the simplest case, where all the variational orbitals are symmetrically equivalent, the Koopmans correction results in a constant shift of the valence and conduction bands, opening the band gap. For systems with non-equivalent variational orbitals the Koopmans correction also affects inter- and intra-band distances, bandwidths, effective masses, \emph{etc.}. This is discussed in more detail in Supplemental Material S2 \cite{SM}.

Third, the computational cost of Koopmans functional calculations is dominated by the calculations to obtain the screening parameters. This procedure is detailed in Supplemental Material S3 \cite{SM}, including a discussion of how these calculations scale. In brief, the calculations to obtain the screening parameters amount to computing the energies of various charged defects. In this context, image charge corrections are especially important (see Supplemental Material S4 \cite{SM}).

The accuracy of Koopmans functionals has already been demonstrated across a range of materials \cite{colonna2019koopmans,de2022bloch,colonna2022koopmans,marrazzo2024spindependent}. For more details about Koopmans functionals, we refer the reader to Ref.~\cite{linscott2023koopmans}.

\subsection{Band alignment with Koopmans functionals}
\label{sec:odd_band_alignment}
\noindent To calculate band alignment with Koopmans functionals, we must adapt the procedure described earlier in Section~\ref{sec:dft_band_alignment}. Here, we take advantage of the aforementioned fact that ground state KI and pKIPZ energies and densities match those of the base functional. This means that when predicting band alignment with Koopmans functionals, the slab calculations only need to be performed at the level of DFT, because --- being a ground-state property --- the average electrostatic potential remains unchanged by the Koopmans correction \cite{stevanovic2014assessing,PhysRevLett.100.186401}. This substantially reduces the cost of this framework for computing band alignment. 

%added in the paper
However, it is important to note that this comes with a caveat. Since the ground-state energies of the base functional and the corrected ones (KI@PBE or pKIPZ@PBE) are identical, this implies that neither PBE, KI@PBE, nor pKIPZ@PBE can reproduce the experimental phase stability of the reported TiO\textsubscript{2} polymorphs. That being said, the third Koopmans functional KIPZ does not share this property and therefore modifies both the ground-state energy and density, and could therefore be used to address relative phase stability more reliably. However, applying KIPZ in the context of this work would require performing all the slab calculations fully at the KIPZ level, significantly increasing the computational cost. 
%%

% However, it is important to note that this comes with a caveat, meaning that since the ground state energies of the base functional and the corrected one (KI@PBE or pKIPZ@PBE) are the same, this also implies that neither PBE nor the above mentioned Koopmans corrected functionals can obtain the experimental phase stability for the reported polymorphs. That being said, it is worth mentioning that deploying the third Koopmans functional (KIPZ) mentioned in 

Proceeding with the KI and pKIPZ functionals, equations \ref{eqn:ip_via_slab} and \ref{eqn:ea_via_slab} become: 
\begin{equation}
\label{eqn:ip_via_slab}
    \mathrm{IP} = \Delta V^\mathrm{DFT} - (\varepsilon_\mathrm{VBM}^\mathrm{DFT}+\Delta \varepsilon_\mathrm{VBM})
\end{equation}
and
\begin{equation}
\label{eqn:ea_via_slab}
    \mathrm{EA} = \Delta V^\mathrm{DFT} - (\varepsilon_\mathrm{CBM}^\mathrm{DFT}+\Delta \varepsilon_\mathrm{CBM})
\end{equation}
where $\Delta \varepsilon_\mathrm{VBM}$ and $\Delta \varepsilon_\mathrm{VBM}$ are the shifts in the bulk band edges due to the Koopmans correction.
% In both of the equations $\Delta V$ depends only on the charge density, moreover, KI correction does not change the total energy of the system which is a function of density, but instead changes the derivatives affecting in that way only spectral properties of materials while preserving the potential energy surface of base DFT functional. This suggests that $\Delta V$ can be solely obtained on a DFT level since the charge density is the ground state property \cite{stevanovic2014assessing,PhysRevLett.100.186401}. And while $\Delta V$ is unchanged  valence band maximum (VBM) and conduction band minimum (CBM) must be updated, hence  we express them as a sum of a self-consistent DFT results for the highest occupied and lowest unoccupied state and a constant shift of Koopmans states with respect to the DFT ones i.e. $VBM_{DFT} +\Delta VBM$; $CBM_{DFT}+ \Delta CBM$ . More precisely the  term in bracket represents the position of Koopmans level relative to the bulk potential.
Equivalently, one can calculate the electron affinity (EA) by simply subtracting the value of electronic gap calculated of Koopmans level from the IP:
\begin{equation}
\label{eqn:ea_via_slab_2}
    \mathrm{EA} = \mathrm{IP} - E_g^\mathrm{Koopmans}
\end{equation}

% In this approach we only need to perform bulk calculations on Koopmans level while slab calculation on a DFT level are enough to provide the bulk potential of the system. As a final result we can level IP and EA on an energy scale with respect to the vacuum level which gives us as a result the band alignment  of a given material. 

\section{Method}\label{Method}

\noindent In this work, we present the band gaps and band alignment of three common polymorphs of TiO\textsubscript{2}. The crystal structure of these three polymorphs were obtained from the Materials Cloud three-dimensional crystal database (MC3D) \cite{MC3D}, which are structures whose geometries have been optimized using semi-local DFT. In the case of anatase and brookite, the optimized lattice parameters are $a$ = $b$ = 3.8001~\AA, and $c$ = 9.6814~\AA\ for anatase and $a$ = 5.1689~\AA, $b$ = 9.2548~\AA, and $c$ = 5.5035~\AA\ for brookite. They differ by less than 2\% from experimentally-reported values ($a$ = $b$ = 3.7842~\AA\ and $c$ = 9.5146~\AA\ for anatase and $a$ = 5.13~\AA, $b$ = 9.16~\AA, and $c$ = 5.43~\AA\ for brookite). For rutile, the optimized lattice parameters are $a$ = 4.60~\AA\ and $c$ = 2.96~\AA\ (which is in excellent agreement with the experimental $a$ = $b$ = 4.5937~\AA\ and $c$ = 2.9581~\AA). There are many reported studies, both experimental and theoretical, regarding the crystal structure of TiO\textsubscript{2} \cite{isaak1998elasticity, asahi2000electronic, dima2021electronic, banerjee2019quantitative, burdett1987structural, mashimo2017structure,gateshki2007titania}.

Koopmans functionals calculations were performed using \texttt{Quantum ESPRESSO} via the \texttt{koopmans} package \cite{linscott2023koopmans,giannozzi2009quantum,Giannozzi_2017}. The screening parameters were calculated via finite-differences, which necessitate the use of a supercell and image charge correction schemes to avoid spurious interactions between periodic images \cite{10.1063/1.477923} (For more details, see Supplemental Material S4 \cite{SM}). For these calculations, we used a $2\times2\times2$ supercell for rutile and anatase and $2\times1\times2$ supercell for brookite. To convert these $\Gamma$-only results to a well-sampled primitive cell the eigenvalues are unfolded to the equivalent primitive cell \cite{de2022bloch}. We used norm-conserving pseudopotentials from the SG15 library (version 1.0) \cite{scherpelz2016implementation}), and a wave-function energy cutoff of 80 Ry as recommended by force convergence data. All of the input and output files can be found on the Materials Cloud Archive  \cite{TiO2MaterialsCloudArchive}.

% To obtain the average macroscopic electrostatic potential ($\Delta V$), we first calculate the variation of the electrostatic potential as a planar average of the electrostatic potential: \begin{equation}
%      \overline{V} = \frac{1}{S}\int_S dxdy V(r) 
%  \end{equation}
%  where $S$ is the cross-sectional area of a unit cell. This planar averaged potential has periodic oscillations along the $z$ axis due to the spatial distribution of electrons and ionic cores. A way to remove these oscillations is by macroscopically averaging the potential:
%  \begin{equation}
% \overline{\overline{V}}(z) = \frac{1}{L} \int_{-\frac{L}{2}}^{+\frac{L}{2}} \overline{V}(z)dz
%  \end{equation}
% Here we integrate over the length of the period of oscillations along $z$. By definition, a macroscopic averaging procedure should give a constant value of an electrostatic potential inside the slab.

% Moving this figure and table here to make it appear in the right place
\begin{figure*}
    \centering
    \includegraphics[width=\textwidth]{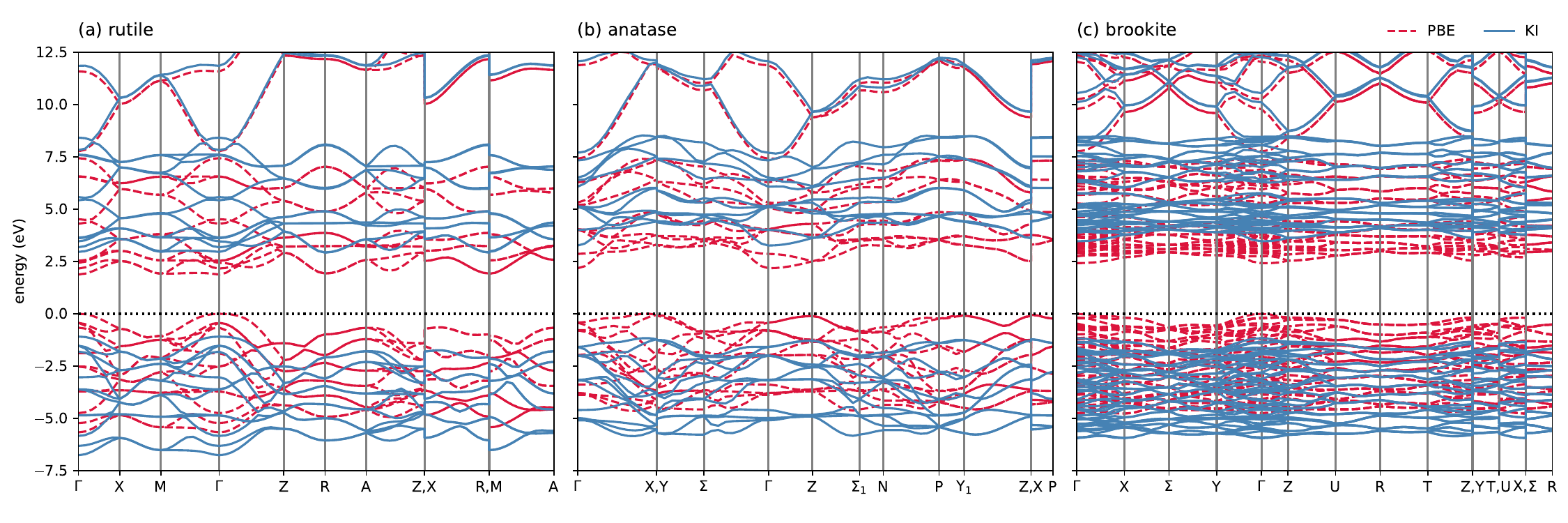}
   \caption{KI@PBE band structures (blue) with respect to the PBE bands (red); the zero reference energy is set to $\varepsilon_\mathrm{VBM}^\mathrm{DFT}$.}
    \label{fig:ki_band_structures}
\end{figure*}

For the slab calculations, we used the thermodynamically most stable facets of $\mathrm{TiO_2}$ polymorphs: the (110) surface facet of 
 rutile, (101) of anatase, and (210) of brookite. These surface orientations have been extensively studied in the literature \cite{thomas2003resonant,wendt2006formation,martinez2011steps,lazzeri2001structure,diebold2003one,hengerer2000structure,ahmed2011photocatalytic,de2014theoretical,hansen2011direct}
. The slab models were constructed using six and eight periodic units for rutile and anatase, and six periodic units for the larger brookite unit cell. 
The depth of the vacuum was chosen to be six times the height of the bulk primitive cell (ensuring correspondence in size between primitive and supercell), which corresponds somewhere between 20 and 30~\AA\ depending on the polymorph.
These slab sizes were chosen to ensure the macroscopic average potential was converged to within 0.01 eV. The slab calculations used a $3\times7\times1$ $k$-point mesh for rutile and anatase and $3\times2\times1$ for brookite. These DFT calculations were also performed using \texttt{Quantum ESPRESSO}. The geometries of the slab structures were optimized using the PBE functional in order to obtain an accurate physical picture of the surface/vacuum interface.%  (As The value of average potential inside the slab can be affected by relaxation\cite{weston2018accurate}.) 

%The necessity of using superlattice calculation is emphasized in literature  \cite{van1987theoretical}, and the main reason for it is the fact that interface calculations are not feasible because of the periodic boundary conditions. Therefore we need to make sure that the material in the superlattice is large enough to contain a bulk-like region.

\section{Results and discussion}\label{Results and discussion}
\subsection{Band structure}

\noindent In Figure~\ref{fig:ki_band_structures} we present the band structures of three polymorphs of TiO\textsubscript{2} calculated via the KI functional, and compare them against those calculated with PBE (the underlying base functional). The band gaps of rutile and brookite TiO\textsubscript{2} are direct ($\Gamma \rightarrow\Gamma$), while that of anatase is indirect ($Z \rightarrow\Gamma$) \cite{deak2011band,zhang2019challenges,D1RA09057G}. In all three cases, the KI correction opens the band gap, shifting the valence bands downwards and the conduction bands upwards. %, leading to the band gap widening relative to the PBE result.  % Knowing the values of a VBM and CBM both on PBE and KI level one calculates the potential shifts $\Delta \varepsilon_\mathrm{VBM}$ and  $\Delta \varepsilon_\mathrm{CBM}$.
(The numerical shifts in the valence band maximum and conduction band minimum are provided in the Supplemental Material \cite{SM}.) While the effect of the Koopmans correction may appear to be a rigid scissors shift, this is not the case; rather, it is an orbital-dependent correction whose effect on individual bands depend on the screening parameters and the orbital character. A full derivation of the KI potential and the conditions under which it reduces to a rigid shift is provided in Supplemental Material S2 \cite{SM}.

The resulting band gaps of the titania polymorphs are presented in Table~\ref{tab:table2}, alongside experimental values, as well as those calculated using hybrid functionals and GW. When comparing against experimental band gaps it is important to account for zero-point renormalization --- by subtracting the ZPR (as calculated \emph{ab initio}) from experimental values one obtains a value that can be fairly compared against the computational results obtained for pristine crystal geometries.

\begin{table}[t]
\caption{Electronic band gaps (in eV) of $\mathrm{TiO_2}$ polymorphs at different levels of theory.}
\label{tab:table2}
\centering
\setlength{\tabcolsep}{5pt}
\begin{tabular}{l *{3}{D{.}{.}{2.3}}}
\toprule
& \multicolumn{1}{c}{rutile}
& \multicolumn{1}{c}{anatase} 
& \multicolumn{1}{c}{brookite} \\
\cmidrule{2-4}
PBE & 1.88 & 2.27 & 2.42 \\
HSE06\footnotetext[1]{\cite{landmann2012electronic}}\footnotemark[1]\footnotetext[2]{\cite{doi:10.1021/ct500087v}}\footnotemark[2] & \multicolumn{1}{c}{3.39, 3.61} &3.60 &3.30 \\
G\textsubscript{0}W\textsubscript{0}@PBE\footnotemark[1]\footnotetext[3]{\cite{kang2015influence}}\footnotemark[3] & 3.66,3.46&4.03,3.73&3.45\\
G\textsubscript{0}W\textsubscript{0}@HSE06\footnotemark[2] & 4.73 & \multicolumn{1}{c}{$-$} & \multicolumn{1}{c}{$-$} \\
GW\textsubscript{0}@PBE\footnotemark[3] & 4.23 & 4.60  & \multicolumn{1}{c}{$-$} \\

%GW@PBE\footnotemark[3] & 4.84 & 5.28 & \multicolumn{1}{c}{$-$} \\

QSGW\footnotetext[4]{\cite{chen2015accurate}}\footnotemark[4]\footnotemark[2]& 4.18, 4.22& \multicolumn{1}{c}{$-$} & \multicolumn{1}{c}{$-$} \\
QSG\~W\footnotetext[5]{\cite{cunningham2023qs}}\footnotemark[4]\footnotemark[5]& 3.73,3.88& \multicolumn{1}{c}{$-$} & \multicolumn{1}{c}{$-$} \\
%QSG\~W\footnotetext[5]{\cite{cunningham2023qs}}\footnotemark[5]& 3.88& \multicolumn{1}{c}{$-$} & \multicolumn{1}{c}{$-$} \\
%G\textsubscript{0}W\textsubscript{0}@HSE06\footnotemark[2] & 4.73 & \multicolumn{1}{c}{$-$} & \multicolumn{1}{c}{$-$} \\
%scQPGW@HSE06\footnotemark[2] & 4.18 & \multicolumn{1}{c}{$-$} & \multicolumn{1}{c}{$-$} \\
GW\footnotemark[3] & 4.84 & 5.28 & \multicolumn{1}{c}{$-$} \\

KI@PBE & 4.04& 4.51& 4.63\\
pKIPZ@PBE & 4.19&4.64 &4.78 \\
exp $-$ ZPR & \multicolumn{1}{c}{$3.1-4.1$} & \multicolumn{1}{c}{$3.4-3.9$} &\multicolumn{1}{c}{$-$}\\ 
exp\footnotemark[6]\footnotetext[6]{\cite{C4CP02201G,xiong2007photoemission,ma14143918,tezuka1994photoemission,rangan2010energy,pascual1978fine,amor1997structural,eufinger2007photocatalytic,kavan1996electrochemical,koelsch2002comparison,li2007anatase,reyes2008phase,hu2009high}}& \multicolumn{1}{c}{$2.8-3.8$} & \multicolumn{1}{c}{$3.2-3.7$} &\multicolumn{1}{c}{$3.1-3.5$}\\ 
& -0.337, \\
ZPR\footnotemark[7]\footnotetext[7]{\cite{miglio2020predominance,engel2022zero,wu2020theoretical}} & -0.349, & -0.238 & \multicolumn{1}{c}{$-$} \\
& -0.314 \\
%KI$-$ZPR& \multicolumn{1}{c}{$2.94-2.98$} & 3.56 & \multicolumn{1}{c}{$-$} \\
%pKIPZ$-$ZPR& \multicolumn{1}{c}{$3.08-3.12$} & 3.71 & \multicolumn{1}{c}{$-$} \\
\bottomrule
\end{tabular}
\end{table}

For rutile and anatase, the electronic band gaps calculated using KI@PBE are within 0.36~eV of experiment (when accounting for ZPR). This level of accuracy is a drastic improvement on that of the base functional PBE, is comparable to that of G\textsubscript{0}W\textsubscript{0}@PBE and hybrid functionals, and is better than G\textsubscript{0}W\textsubscript{0}@HSE06 and self-consistent quasiparticle GW. For brookite, no ZPR correction was found in the literature and thus comparison with experiment is not possible.

%\begin{figure*}[t]  
%\centering
%\begin{subfigure}[H]{0.32\textwidth}
%\includegraphics[width=\linewidth]{Figures/pKIPZrutile2.png}   \caption{Rutile}
%\end{subfigure}
%\begin{subfigure}[H]{0.32\textwidth}
%%\includegraphics[width=\linewidth]{Figures/pKIPZanatase2.png}    
%\caption{Anatase}    
%\end{subfigure}
%\begin{subfigure}[H]{0.32\textwidth}
%\includegraphics[width=\linewidth]{Figures/pKIPZbrookite2.png} 
%\caption{Brookite}
%\end{subfigure}
%%\caption{pKIPZ band structures with respect to the PBE bands, with the reference energy set to $\varepsilon_\mathrm{VBM}^\mathrm{DFT}$.}
%\label{fig:pkipz_band_structures}
%\end{figure*}
\newpage

The results with pKIPZ are very similar to that of KI (plots of pKIPZ@PBE band structures can be found in the Supplemental Material \cite{SM}). Compared to the KI functional, pKIPZ reports slightly larger band gaps, especially in the case of brookite. Both KI and pKIPZ assign brookite the largest band gap of the three polymorphs, in contrast to HSE06 and G\textsubscript{0}W\textsubscript{0}@PBE which assign it the smallest.
%adding a part about HSE06 performance
For these systems, HSE06 reproduces the experimental band gap more accurately than KI@PBE and pKIPZ@PBE. This difference can be understood in light of the distinct physical principles underlying hybrid and Koopmans functionals. Hybrid functionals incorporate a fraction of exact exchange and often achieve reasonable band gaps through a fortuitous cancellation of errors between semi-local DFT—whose convexity error leads to underestimated gaps—and Hartree–Fock, which overestimates gaps due to concavity. Since the HSE06 mixing parameter was calibrated using molecular atomization energies \cite{heyd2003hybrid}, its improved agreement with experiment here is likely incidental rather than systematic.
Koopmans functionals, by contrast, impose generalized piecewise linearity to improve spectral properties, and their accuracy depends critically on the quality of the 
$\Delta$SCF energies produced by the base functional. Thus, the superior performance of HSE06 in this case may reflect particularly favorable error cancellation in HSE06 or limitations in the $\Delta$SCF accuracy of PBE. (We note that KI's overestimation of the band gap is consistent with earlier work, and is under ongoing investigation \cite{nguyen2018koopmans}.) Using a more accurate base functional could improve Koopmans predictions (at the expense of increased computational cost), although this remains to be explored.

The other notable difference between hybrid and Koopmans functionals is their computational cost. The computational cost of Koopmans functional calculations is dominated by the evaluation of the screening parameters, which can be obtained either through a $\Delta$SCF approach based on charged supercell calculations or through a more efficient DFPT reformulation in the primitive cell \cite{nguyen2018koopmans,de2022bloch,linscott2023koopmans,colonna2022koopmans,Colonna2018}.  In this work we employ the $\Delta$SCF method, which requires supercells and charge corrections and whose cost depends on the number of symmetry-inequivalent variational orbitals. A detailed discussion of both approaches and their scaling is provided in the Supplemental Material S3 \cite{SM}. For comparison, we note that hybrid functionals typically scale as $\mathcal{O}(N^4)$ due to the non-local exchange term~\cite{laqua2018efficient}.

%added part
Table~\ref{tab:table2} shows that the KI and KIPZ band gaps for rutile are in closer agreement with experiment than those for anatase and brookite, and that the differences among the three TiO$_2$ polymorphs are more pronounced with KI and KIPZ than with HSE06 or $G_0W_0$. At present, no definitive explanation for this behavior can be offered. Given the wide spread in the reported experimental band gaps and the limited number of available data-points, drawing firm conclusions would be premature. There is no fundamental reason to expect differences between polymorphs to be inherently more pronounced when using Koopmans functionals than when using hybrid functionals or GW. Ongoing comprehensive benchmark studies are expected to clarify such trends more quantitatively.

%But
%However
Finally, without an accurate measurement of the ZPR in brookite and more precise experimental measurements of the band gap, it is difficult to determine which ordering is correct.

\subsection{Band alignment}
\begin{table*}[]
\caption{Ionization potentials  and electron affinities in eV obtained on KI and pKIPZ level in comparison with experimental studies of single-crystal polymorphs. \label{tab:table3}}
\centering
\begin{tabular}{l *{2}{D{.}{.}{3.4}} @{\extracolsep{0.5em}} *{2}{D{.}{.}{3.4}} @{\extracolsep{0.5em}} *{2}{D{.}{.}{3.4}}}\toprule 
 &\multicolumn{2}{c}{rutile (110)} &\multicolumn{2}{c}{anatase (101)} &\multicolumn{2}{c}{brookite (210)}\\
  \cmidrule{2-3}
  \cmidrule{4-5}
  \cmidrule{6-7}
  & \multicolumn{1}{c}{IP} &
  \multicolumn{1}{c}{EA} &
  \multicolumn{1}{c}{IP} &
  \multicolumn{1}{c}{EA} &
  \multicolumn{1}{c}{IP} &
  \multicolumn{1}{c}{EA} \\
  PBE &  7.29 & 5.41 & 7.43 & 5.16 & 7.17 & 4.75\\
  HSE06\footnotetext{\cite{doi:10.1021/ct500087v}}\footnotemark[1] & 8.66 & 4.99 & \multicolumn{1}{c}{$-$}  & \multicolumn{1}{c}{$-$}  & \multicolumn{1}{c}{$-$}  & \multicolumn{1}{c}{$-$} \\
  HSE06 (QM/MM)\footnote{\cite{scanlon2013band}} & 7.83 & \multicolumn{1}{c}{$-$}  & 8.3 & \multicolumn{1}{c}{$-$}  & \multicolumn{1}{c}{$-$}  & \multicolumn{1}{c}{$-$} \\
  G\textsubscript{0}W\textsubscript{0}@PBE\footnotemark[1] & 7.29 & 3.03 & \multicolumn{1}{c}{$-$}  & \multicolumn{1}{c}{$-$}  & \multicolumn{1}{c}{$-$}  & \multicolumn{1}{c}{$-$}  \\
  G\textsubscript{0}W\textsubscript{0}@HSE06\footnotemark[1] & 7.92 & 3.19 & \multicolumn{1}{c}{$-$}  & \multicolumn{1}{c}{$-$}  & \multicolumn{1}{c}{$-$}  & \multicolumn{1}{c}{$-$}  \\
  scQPGW@HSE06\footnotemark[1] & 8.77 & 3.59 & \multicolumn{1}{c}{$-$}  & \multicolumn{1}{c}{$-$}  & \multicolumn{1}{c}{$-$}  & \multicolumn{1}{c}{$-$}  \\
  KI@PBE & 8.38 & 4.34 & 8.59 & 4.08 & 8.34 & 3.71 \\ 
  pKIPZ@PBE & 8.14 & 3.95 & 8.33 & 3.69 & 8.08 & 3.30 \\
  exp $-$ ZPR & 8.5 & \multicolumn{1}{c}{$4.97 - 5.00$} & \multicolumn{1}{c}{$-$}  & \multicolumn{1}{c}{$-$}  & \multicolumn{1}{c}{$-$}  & \multicolumn{1}{c}{$-$}  \\
exp\footnote{\cite{thomas2003resonant,kashiwaya2018work,thomas2007comparison}} & 8.20 &\multicolumn{1}{c}{$5.14 - 5.17$}   & \multicolumn{1}{c}{7.96, 8.07, 8.20} & \multicolumn{1}{c}{$-$} & \multicolumn{1}{c}{$-$}  & \multicolumn{1}{c}{$-$}   \\
  ZPR\footnote{\cite{deMelo2023}} & -0.3 & 0.17 & \multicolumn{1}{c}{$-$} & \multicolumn{1}{c}{$-$} & \multicolumn{1}{c}{$-$} & \multicolumn{1}{c}{$-$}\\
%  \multicolumn{1}{c}{exp.\footnote{\cite{thomas2003resonant,kashiwaya2018work,nguyen2018koopmans,thomas2007comparison}}} &
%  \multicolumn{1}{c}{KI} &
%  \multicolumn{1}{c}{pKIPZ}\\
%\cmidrule{2-4}
%\cmidrule{5-6}
%rutile (110) & 8.00  &7.85 &8.20 &4.71 &4.42 \\
%anatase (101) & 7.91  &8.04 & \multicolumn{1}{c}{7.96, 8.07, 8.20} &4.11 &4.09 \\
%brookite (210) & 8.03 &7.86 &\multicolumn{1}{c}{$-$} &4.03&3.69  \\
\bottomrule
\end{tabular}
\end{table*}

\noindent The ionization potentials and electron affinities of the $\mathrm{TiO_2}$ polymorphs, as calculated via KI and pKIPZ following the method described in Section~\ref{sec:odd_band_alignment}, are reported in Table~\ref{tab:table3} alongside values obtained from other higher levels of theory and experiment. Once again it is important to account for ZPR when comparing against experiment. However, ZPR shifts for the VBM and CBM individually have only been reported for the rutile polymorph, calculated using a generalized Fr\"ohlich model~\cite{deMelo2023}. This technique is more approximate than those that were used to calculate the ZPR shifts in the band gaps listed in Table~\ref{tab:table2}. Note that the combining the shifts for the VBM and CBM predicted by the Fr\"ohlich model results in a ZPR reduction of the band gap (0.47 eV) that is up to 50\% larger than those listed in Table~\ref{tab:table2}, so these results should be treated with caution. That caveat aside, for rutile KI@PBE underestimates the shifted experimental IP by 0.5 eV --- comparable to the accuracy of G\textsubscript{0}W\textsubscript{0}@HSE06. The EA, on the other hand, is predicted by KI@PBE within the range of values reported experimentally --- a marked improvement upon the GW results, and matched only by HSE06. For anatase and brookite, no ZPR results for the IP and EA were found. The KI and pKIPZ result for the IP of anatase are fractionally lower than that given by HSE06 \cite{scanlon2013band}.

\begin{figure}
    \includegraphics[width=\columnwidth]{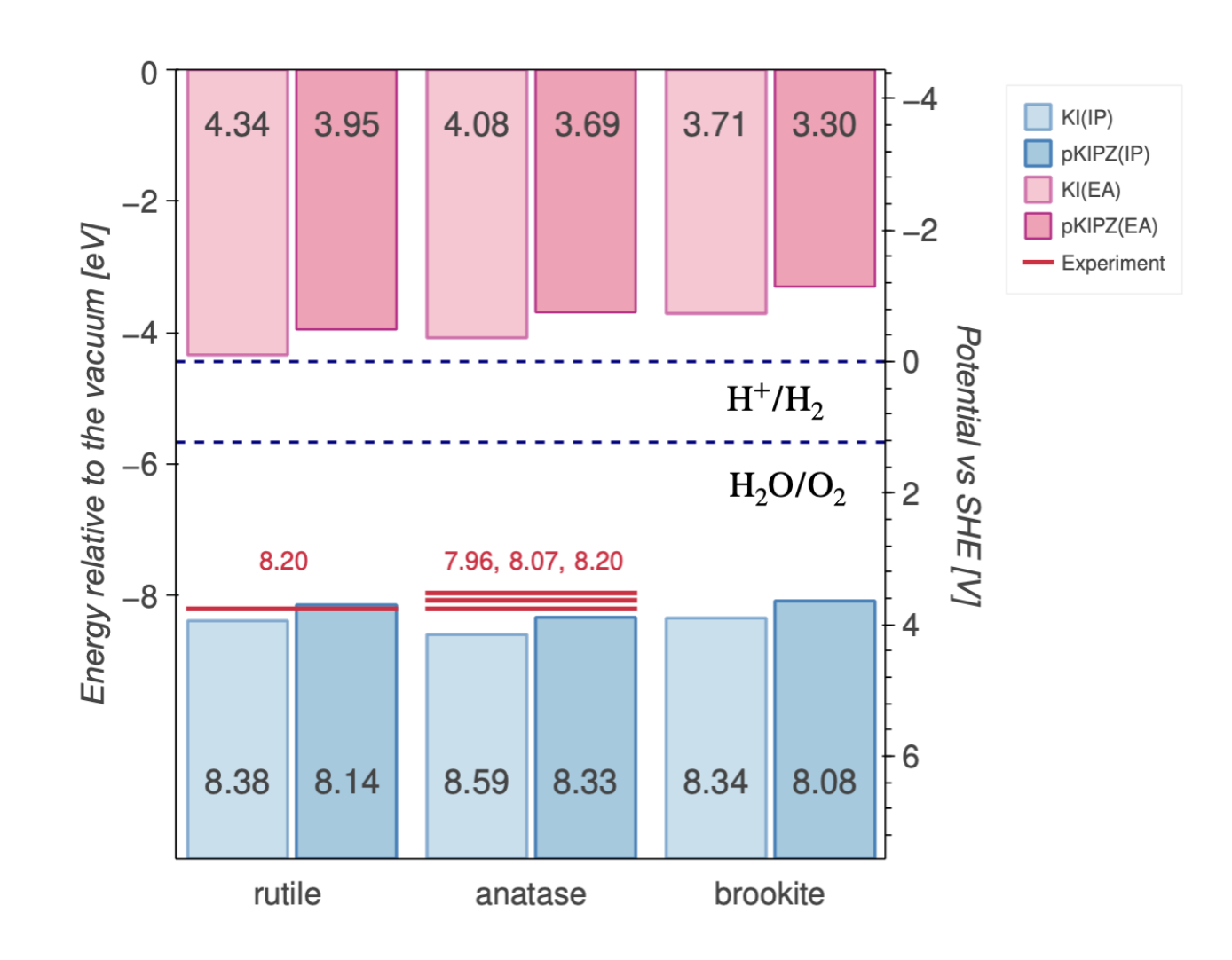}
    \caption{Band alignment of TiO$_2$ polymorphs using KI and pKIPZ. Blue rectangles represent the IPs of three polymorphs, while red rectangles the EAs. Experimental values are given as red solid lines (from Refs.~\cite{thomas2003resonant,kashiwaya2018work,nguyen2018koopmans,thomas2007comparison}). }
    \label{fig:mesh4}
\end{figure}

The band alignment diagram represented in Figure \ref{fig:mesh4} shows the IPs and EAs of the three polymorphs compared against the two redox reaction potentials. The H\textsuperscript{+}/H\textsubscript{2} potential lies at -4.44~eV \cite{trasatti1986absolute}, while the H\textsubscript{2}O/O\textsubscript{2} potential is 1.23~eV below this (as per Equation~\ref{eqn:H2O_ox}) \cite{CRCHandbook}. The positions of the valence and conduction bands should straddle these two potentials i.e.\ the VBM and CBM should not lie within the two dashed lines in Figure~\ref{fig:mesh4}. According to our calculations, rutile TiO\textsubscript{2} appears to satisfy this condition, but only marginally, as its conduction-band minimum lies only slightly above the reduction potential. This borderline alignment may help explain why—despite having a more desirable band gap than anatase and brookite—rutile often performs as a less efficient photocatalyst, particularly relative to anatase\cite{luttrell2014anatase}.
% appears not to satisfy this condition, which may explain why --- despite having a more desirable band gap than anatase and brookite --- this polymorph is a slightly worse catalyst, especially compared to anatase \cite{luttrell2014anatase}. 
Brookite, on the other hand, shows some promise as a photocatalyst: the positions of the valence and the conduction bands seem to be favorable, but its fractionally larger band gap would limit the amount of photons that would be absorbed when subjected to sunlight. In this case, one might be able to engineer the band gap.
%\begin{figure}[t]
%    \centering
%    \includegraphics[width=\columnwidth]{Figures/BA.svg}
%    \caption{Band alignment of TiO$_2$ polymorphs using pKIPZ functional. Blue rectangles represent the IPs of three polymorphs, while red rectangles show EAs. Experimental values are given in red solid lines from references listed in Table\ref{tab:table3} }
%    \label{fig:mesh5}
%\end{figure}
The KI and pKIPZ results are qualitatively the same, with pKIPZ reporting a slightly more favorable CBM for rutile relative to the hydrogen reduction reaction. Still, with respect to the value of band gap and band edge positions, both KI and pKIPZ predict that anatase appears to be the most promising photocatalyst of the three.

Of course, an optimal band alignment does not guarantee photocatalytic success: anatase also exhibits higher surface activity and desirable excitonic properties \cite{luttrell2014anatase,eidsvaag2021TiO2}. 
%% added part(referee 1. question 2)
Indeed, excitonic binding energies and lifetimes vary significantly across the polymorphs. Rutile hosts weakly bound excitons (4–26 meV), with experiments reporting a 4 meV exciton \cite{pascual1977resolved} and GW calculations placing the lowest dipole-allowed exciton at 26 meV \cite{kang2010quasiparticle}; these small binding energies allow thermal dissociation, and charge carriers in rutile exhibit very short lifetimes \cite{yamada2012determination}. In contrast, anatase supports more strongly bound excitons, with GW calculations predicting a 160 meV exciton in the (001) plane \cite{baldini2017strongly}, and a recent study including lattice distortions reporting exciton–polaron binding energies of 216 meV \cite{dai2024identification}. Experiments also indicate that charge carriers in anatase are substantially longer-lived than in rutile \cite{yamada2012determination}. (To our knowledge, excitonic data for brookite are not available.) Importantly, these excitonic effects are not strong enough to appreciably shift the band-edge positions relative to the redox potentials and therefore do not influence the band-alignment results presented here.

\section{Conclusions}

\noindent In conclusion, this investigation demonstrates the accuracy of Koopmans spectral functionals for calculating the band gap, IP, and EAs of TiO\textsubscript{2}, finding --- in agreement with experiment --- that anatase is the most promising photocatalyst of the three. The individual predictions of the ionization potentials and electron affinities were either as or more accurate than the results of hybrid functionals and many-body perturbation theory. At the same time, the band alignment calculations presented in this work are notably simpler, given that this framework only requires Koopmans functional calculations on the primitive cell and semi-local DFT calculations for the slab calculations. In the future, Koopmans functionals could be deployed to screen promising photocatalyst candidates that are much less well-studied. 

\section{Acknowledgments}
\noindent MS acknowledges support from EXAF (Excellence in Africa) Research Centre as part of the collaboration ``Embedded exact quantum dynamics for photocatalytic water splitting". This research was supported by the NCCR MARVEL, a National Centre of Competence in Research, funded by the Swiss National Science Foundation (grant number 205602).
%\nocite{*}
%\bibliographystyle{plain}
%\nocite{*}
\bibliography{tio2}%\bibliography{apssamp}% Produces the bibliography via BibTeX.⁄

\end{document}

\title{Supplemental Material for: Predicting the suitability of photocatalysts for water splitting using Koopmans spectral functionals: The case of TiO\textsubscript{2} polymorphs} %Force line breaks with \\

\author{Marija Stojkovic}
\email{marija.stojkovic@epfl.ch}
\affiliation{Theory and Simulations of Materials (THEOS), \'{E}cole Polytechnique F\'{e}d\'{e}rale de Lausanne, 1015 Lausanne, Switzerland}
\author{Edward Linscott}
\affiliation{Center for Scientific Computing, Theory and Data, Paul Scherrer Institute, 5232 Villigen PSI, Switzerland}
\affiliation{National Centre for Computational Design and Discovery of Novel Materials (MARVEL), Paul Scherrer Institute, 5232 Villigen PSI, Switzerland}

\author{Nicola Marzari}

\affiliation{Theory and Simulations of Materials (THEOS), \'{E}cole Polytechnique F\'{e}d\'{e}rale de Lausanne, 1015 Lausanne, Switzerland}
\affiliation{Center for Scientific Computing, Theory and Data, Paul Scherrer Institute, 5232 Villigen PSI, Switzerland}
\affiliation{National Centre for Computational Design and Discovery of Novel Materials (MARVEL), Paul Scherrer Institute, 5232 Villigen PSI, Switzerland}

\date{\today}% It is always \today, today,

 %\begin{description}
%\end{description}                           %display desired

\maketitle

\begin{table}[h]
\caption{{VBM and CBM of three polymorphs of TiO\textsubscript{2} obtained using semi-local DFT (PBE) and two flavors of Koopmans functionals (KI and pKIPZ). All values are given in units of eV.} 
\label{tab:B_Table1}}
 \setlength{\tabcolsep}{0.5pt} % Default value: 6pt
 \renewcommand{\arraystretch}{0.55} % Default value: 1
\centering
\footnotesize
\begin{tabular}{ l @{\extracolsep{0.5em}} *{10}{D{.}{.}{3.4}}}
\toprule

&\multicolumn{3}{c}{HOMO} &\multicolumn{3}{c}{LUMO} &\multicolumn{2}{c}{$\Delta VBM$} & \multicolumn{2}{c}{$\Delta CBM$}\\
\toprule
& \multicolumn{1}{c}{PBE}
& \multicolumn{1}{c}{KI}
& \multicolumn{1}{c}{pKIPZ}
& \multicolumn{1}{c}{PBE}
& \multicolumn{1}{c}{KI}
& \multicolumn{1}{c}{pKIPZ}
& \multicolumn{1}{c}{KI}
& \multicolumn{1}{c}{pKIPZ}
& \multicolumn{1}{c}{KI}
& \multicolumn{1}{c}{pKIPZ} \\
\cmidrule{2-4}
\cmidrule{5-7}
\cmidrule{8-9}
\cmidrule{10-11}
rutile & -0.62 & -1.71 & -1.47 & 1.26 & 2.32 & 2.71 & -1.09 & -0.85 &1.06 & 1.46\\
anatase & -1.70 & -2.86 & -2.60 & 0.57 & 1.65 & 2.05 & -1.16 & -0.90 &1.08 &1.48\\
brookite & -1.17 & -2.34 & -2.08 & 1.25 & 2.29 & 2.70 & -1.17 & -0.91 &1.04 &1.45\\
\bottomrule
\end{tabular}
\end{table}
\begin{figure}
    \centering
    \includegraphics[width=\columnwidth]{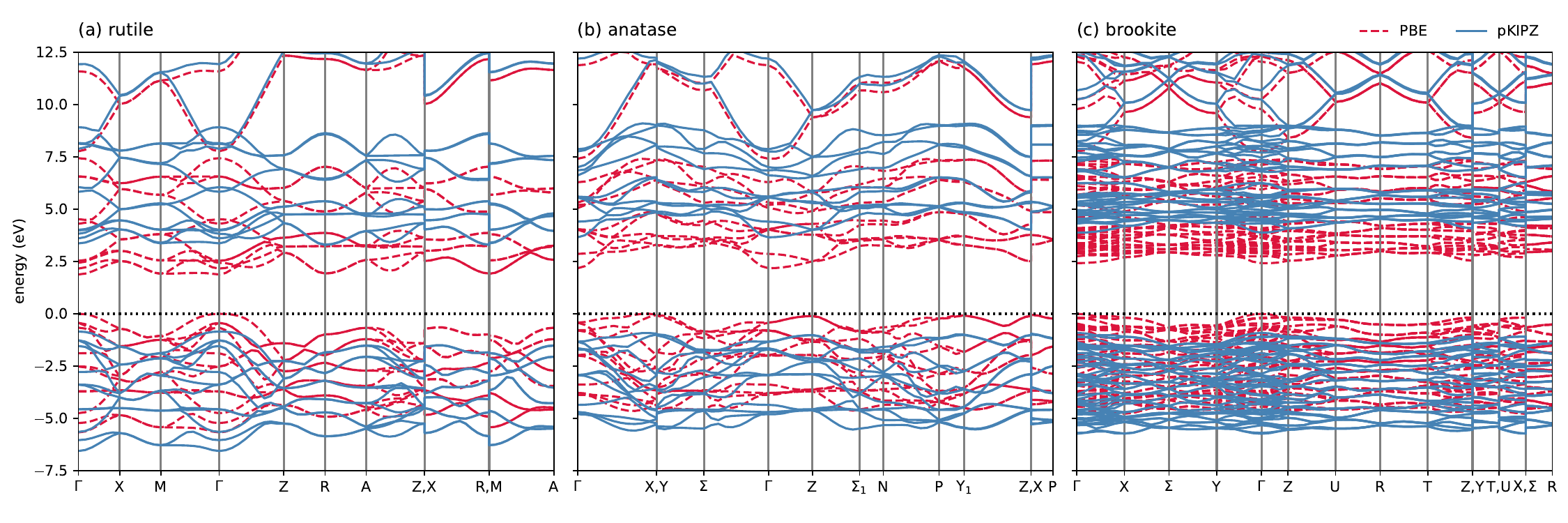}
   \caption{pKIPZ band structures of three polymorphs of TiO\textsubscript{2}, alongside the PBE bands, with the reference energy set to $\varepsilon_\mathrm{VBM}^\mathrm{DFT}$.}
    \label{fig:pkipz_band_structures}
\end{figure} 

\section{Variational calculations with ODD functionals}

\noindent In terms of variational calculations, the orbital-density dependent (ODD) nature of Koopmans functionals makes these calculations more complex than DFT. DFT functionals depend solely on the total density, which means that any set of occupied orbitals that are related via a unitary rotation must yield the same energy. This is not the case for ODD functionals, which break unitary invariance and generally give different energies for different orbital densities (even if the total density is unchanged), and thus we must go beyond standard DFT minimization procedures. 
The variation of $E^\mathrm{Koopmans}$ (equation 8 of the manuscript) with respect to an arbitrary change of each orbital
$\varphi_i$ leads to the Euler-Lagrange equations:
\begin{equation}
    h^{DFT} |\varphi_i \rangle + v_i^{ODD}|\varphi_i \rangle = \sum_j \Lambda_{ij}|\varphi_j \rangle
\end{equation}

 where first term in the equation on the left-hand side is the Hamiltonian of the underlying DFT functional, the second term refers to orbital-density dependent potential associated with the orbital, and on the right-hand side we have Lagrange multipliers that enforce orthonormality. 
 At the minimum, variation with respect to infinitesimal unitary transformations among the occupied orbitals must vanish. This gives the Pederson condition:
 \begin{equation}
     \langle \varphi_ih_i|\varphi_j\rangle = \langle \varphi_i|h_j\varphi_j\rangle  \end{equation}

 The self-consistent solution of these equations define the proper minimum of the Koopmans functionals.
 
 Practically, these equations are solved using a similar approach to that employed in ensemble density functional theory~\cite{marzari1997ensemble}: the ODD energy is minimized directly via two nested loops: an inner loop minimizes the ODD energy with respect to unitary rotations of the variational orbitals (i.e. the total density remains fixed), while an outer loop permits changes in the total density. Both loops use the conjugate gradient algorithm. For more details, see Reference~\cite{borghi2015variational}.
 Finally, orbital occupancies are not treated as variational parameters. While derivatives with respect to orbital occupations are at the core of the Koopmans functional formalism (such as equation 8), in practical calculations we deal exclusively with systems with a non-vanishing band gap and therefore the occupied orbitals always have $f_i = 1$ and the empty orbitals are always $f_i = 0$.

\section{Koopmans correction: Beyond a scissor shift}

\noindent The KI potential is not a rigid shift of the PBE bands (i.e. a ``scissors operator'').
The Koopmans orbital-dependent potential is given by
\begin{equation}
    v_i^\mathrm{KI}(\mathbf{r}) = \frac{\delta E^\mathrm{KI}}{\delta \rho_j(\mathbf{r})} = v^\mathrm{KS}(\mathbf{r}) + \sum_j \alpha_j v^\mathrm{corr}_{ji}(\mathbf{r})
\end{equation}
where the corrective Koopmans potential is given by the sum of three terms:
\begin{equation}
    v^\mathrm{corr}_{ij}(\mathbf{r}) = \delta_{ij} v^\mathrm{scalar}_j + \delta_{ij} v_j^\mathrm{diag}(\mathbf{r}) + (1 - \delta_{ij}) v_j^\mathrm{off\-diag}(\mathbf{r})
\end{equation}
These terms are, respectively, a scalar contribution:
\begin{equation}
  v_j^\mathrm{scalar} = - E_\mathrm{Hxc}[\rho - \rho_j] + E_\mathrm{Hxc}[\rho - \rho_j + n_j] - \int v_\mathrm{Hxc}[\rho - \rho_j + n_j](\mathbf{r}')n_i(\mathbf{r}') \, d\mathbf{r}'
\end{equation}
a real-space but diagonal potential
\begin{equation}
   v_j^\mathrm{diag}(\mathbf{r}) = - v_\mathrm{Hxc}[\rho](\mathbf{r}) + v_\mathrm{Hxc}[\rho - \rho_j + n_j](\mathbf{r})
\end{equation}
and a real-space and off-diagonal correction
\begin{equation}
    v_j^\mathrm{off\-diag}(\mathbf{r}) = (1 - f_j) v_\mathrm{Hxc}[\rho - \rho_j] - v_\mathrm{Hxc}[\rho](\mathbf{r}) + f_j v_\mathrm{Hxc}[\rho - \rho_j + n_j](\mathbf{r})
\end{equation}
where $f_i n_i(\mathbf{r}) = \rho_i(\mathbf{r})$ is the occupation-weighted density of orbital $i$.

This correction clearly amounts to more than a scissors-operator (which would be an occupation-dependent scalar potential). However, in a few cases the Koopmans potential simplifies. To begin, for systems where the orbital occupations are all integer (i.e. 0 or 1; as in the case for insulators), then the terms in $v_j^\mathrm{off\-diag}(\mathbf{r})$ cancel and the correction becomes diagonal. Furthermore, for occupied orbitals, the terms in $v_j^\mathrm{diag}(\mathbf{r})$ also cancel, leaving us with a scalar orbital-dependent potential.

This brings us to the second important point, which is that the Koopmans potential is different for each \emph{variational} orbital, while the band structure corresponds to the eigenvalues of the \emph{canonical} orbitals. For a system with non-uniform screening parameters, the effect of the Koopmans correction on the DFT band structure is more complex, becoming a linear mix of the screening parameters of variational orbitals that constitute the canonical orbital in question. More concretely, given that the variational and canonical orbitals are related via a unitary rotation (i.e. $|\psi_i\rangle = \sum_j U_{ij} |\varphi_j\rangle)$, it follows that the Koopmans correction shifts DFT quasi-particle energies by

\begin{align}
    \Delta \varepsilon_i = \varepsilon^\mathrm{KI}_i - \varepsilon^\mathrm{DFT}_i
    =
    \sum_{jk} \alpha_j U_{ij}U_{ki}^\dag\langle\varphi_k|\hat v_j^\mathrm{KI}|\varphi_j\rangle
\end{align}

\noindent which is proportional to $\alpha_j$, with a constant of proportionality corresponding to the degree of overlap between canonical-variational orbital pairs, as well as Koopmans potential matrix elements. As discussed above, for fully-occupied orbitals the matrix element $\langle\varphi_k|\hat v_j^\mathrm{KI}|\varphi_j\rangle$ is is diagonal and scalar, and the above expression simplifies to
\begin{align}
    \Delta \varepsilon_{i \in \mathrm{occ}} = 
    \sum_{j} \alpha_j U_{ij}U_{ji}^\dag
    \bigg( & -E_\mathrm{Hxc}[\rho - n_j]+E_\mathrm{Hxc}[\rho] - \int d\mathbf{r} \, v_\mathrm{Hxc}[\rho](\mathbf{r})  n_j(\mathbf{r}) \bigg)
\end{align}c
which, if (a) all the screening parameters are the same and (b) all of the variational orbitals are rotationally equivalent, then the shift is the same for all occupied eigenvalues \emph{i.e.} we have a rigid shift.

Of course, all of these simplifying conditions rarely hold. For example, in ZnO the valence bands are comprised of variational orbitals of different character, each of which is subject to a different KI correction, and the resulting KI bands are not just rigidly shifted: there are clear changes in the bandwidth as well as the position of bands \emph{relative to the valence band maximum}. This illustrates that a rigid shift is a special case that emerges only under very specific conditions \cite{linscott2023koopmans}.

\section{Computational cost of Koopmans calculations}

\noindent The computational cost of a Koopmans functional is dominated by calculating the screening parameters. There are two different methods for computing these parameters:
\begin{enumerate}
    \item $\Delta$SCF, where one explicitly computes the energy of charged defects in a supercell~\cite{nguyen2018koopmans,de2022bloch,linscott2023koopmans}
    \item DFPT, where we reformulate the aforementioned problem using $k$-point sampling and linear response theory~\cite{colonna2018screening,colonna2022koopmans}
\end{enumerate}
In this study we used the first method, which consists of calculating all of the energy differences $\Delta E_i^\mathrm{Koopmans}$ via series of constrained Koopmans or DFT calculations. Given an initial guess {$\alpha_i^0$} for the screening parameters, the values for filled orbitals is given by:
\begin{equation}
    \alpha_i^{n+1} = \alpha_i^n \frac{\Delta E_i-\lambda_{ii}^0(1)}{\lambda_{ii}^{\alpha_i^n(1)}-\lambda_{ii}^0(1)}
\end{equation}
while for empty orbitals it is given by
\begin{equation}
   \alpha_i^{n+1} = \alpha_i^n \frac{\Delta E_i - \lambda_{ii}^0(0)}{\lambda_{ii}^{\alpha_i^n}(0) - \lambda_{ii}^0(0)}
\end{equation}
where 
\begin{equation}
    \lambda_{ii}^{\alpha}(f) = \frac{\partial E^\mathrm{Koopmans}} {\partial f_i} \Big |_{f_i=f} = \langle\varphi_i|\hat{H}_{DFT} + \alpha\hat{v}_i^\mathrm{Koopmans}|\varphi_i\rangle \Big |_{f_i=f}
\end{equation}

Since we are dealing with $N$ and $N\pm 1$ systems, supercells and charge corrections are necessary to avoid spurious interactions (as discussed in our answer to point 3). Each of these supercell calculations scales as $\mathcal{O}(N^{SC})^3$, where $N^{SC}$ is the number of electrons in the supercell. The number of these calculations that we must perform is highly dependent on the particular system being studied. For example, a highly-symmetric, ordered crystal all of the variational orbitals might be symmetrically equivalent to one another, so we only need to compute one screening parameter. For more complex systems, the number of symmetrically-unequivalent orbitals will increase. In the case of rutile TiO\textsubscript{2}, there are 24 unique occupied orbitals and 12 unique empty orbitals. 

The scaling of the supercell calculations can be prohibitively expensive, which is why we developed the DFPT strategy. This replaces the explicit charged defect calculations with an equivalent linear response problem, and scales as \[T_{\mathrm{PC}} \propto N_q N_k N_{\mathrm{PC}}^3\]
This is a standard computational time for the SCF cycle ($N_k N_{\mathrm{PC}}^3$), times the number of independent monochromatic perturbations ($N_q$). Using the relation 
\[N_{\mathrm{SC}}=N_kN_{\mathrm{PC}},\]
and the fact that $N_q = N_k$, the ratio between the supercell and primitive cell computational times is: \[\frac{T_{\mathrm{SC}}}{T_{\mathrm{PC}}} \propto N_q.\]
This implies that as the supercell size (and, equivalently, the number of $q$-points in the primitive cell) increases, the primitive cell DFPT approach becomes more computationally convenient. For details of DFPT approach the reader is referred to Ref.~\cite{colonna2022koopmans}.

In comparison, standard hybrid functionals typically scale as $O(N^4)$ due to the non-local nature of the exchange term \cite{laqua2018efficient}

\section{Image charge corrections}

\begin{center}
    \centering
    \includegraphics[width=\textwidth]{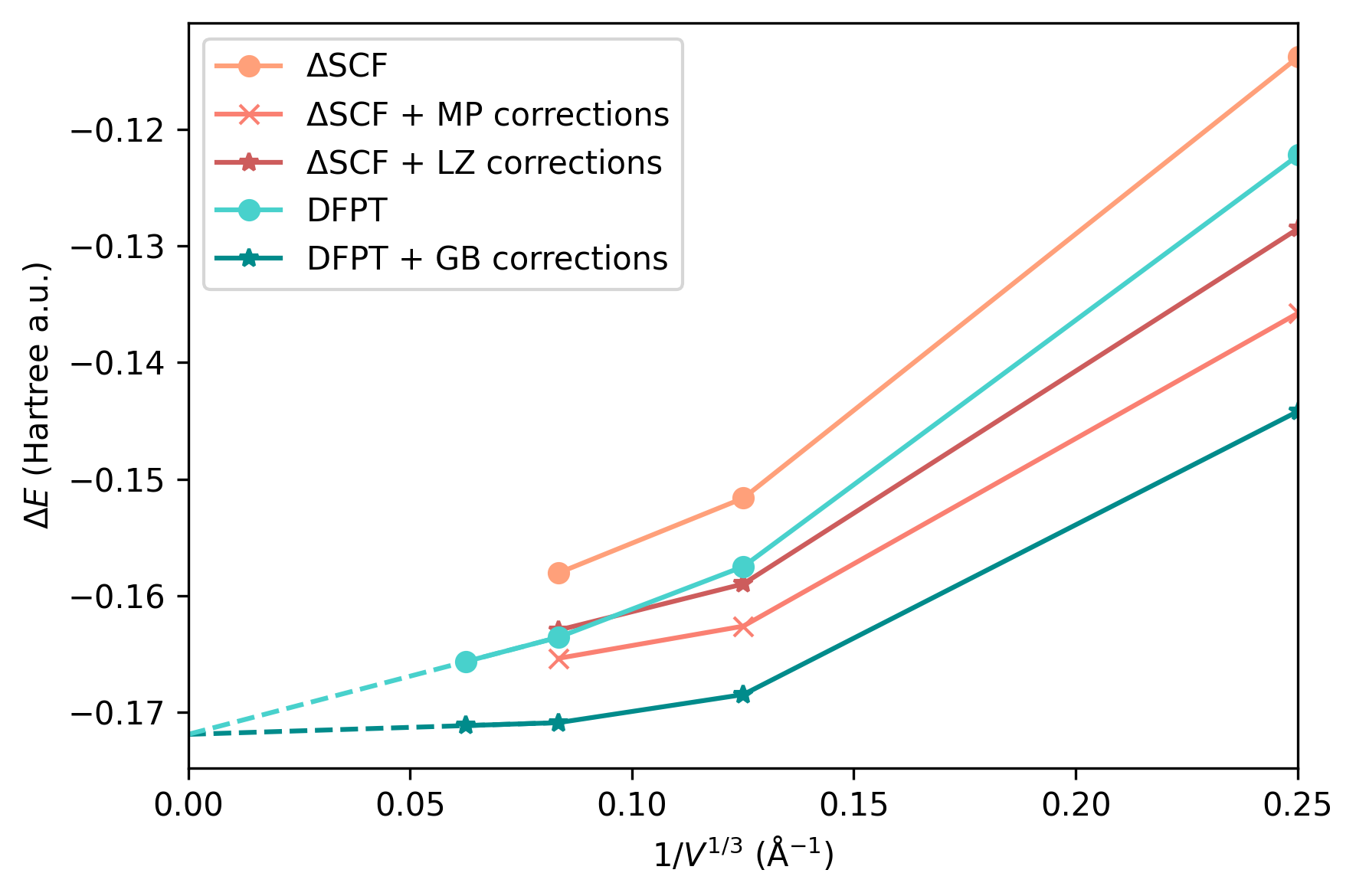}  % Same size for both images
    \hspace{0.5cm}
    \captionof{figure}{The energy difference $\Delta E_i = E_i(N - 1) - E(N)$ from removing one electron from variational orbital $i$ in rutile TiO\textsubscript{2}, as calculated with and without various image corrections schemes --- Makov-Payne (MP), Lany-Zunger (LZ) and Gygi-Baldereschi (GB) --- and using either explicit charged defect calculations ($\Delta$SCF) or density functional perturbation theory (DFPT).}
    \label{fig:charge_corrections}
\end{center}
In the context of Koopmans functionals, image charge corrections are especially relevant when we calculate screening parameters, which involve a series of charged defect (\emph{i.e.} $\Delta$SCF) calculations. To ensure the accuracy of these charged defect calculations, we tested several charge correction schemes for different supercell sizes, starting from $1\times1\times1$ to $4\times4\times4$. These results of these tests are presented in Figure \ref{fig:charge_corrections}.

Focusing firstly on the $\Delta$SCF results (plotted in red), we can see that image charge corrections make the defect energy difference converge much more rapidly. But even with the largest supercell size that we tested ($3\times3\times3$), it was not yet clear if $\Delta E_i$ was converged. In order to obtain results for even larger supercell sizes, we performed additional calculations where the $\Delta$SCF energy is not computed explicitly, but instead computed via density functional perturbation theory (DFPT). This approach scales more favorably as a function of the system size but makes several simplifying approximations. The DFPT results are shown in dark green; we can see that the $3 \times 3 \times 3$ grid is indeed converged, being within 0.2~mHa of the $4 \times 4 \times 4$ result. We therefore conclude that for these calculations, whether $\Delta$SCF or DFPT, we should use a $k$-point grid with a resolution of at least 0.4 \AA\textsuperscript{-1} (\emph{i.e.} that of the $3\times3\times3$ grid) with image charge corrections. This $k$-point resolution equates to $3 \times 3 \times 3$ and $2 \times 1 \times 2$ $k$-point grids for rutile, anatase, and brookite respectively.

\bibliographystyle{unsrt}
\bibliography{reference.bib}